\documentclass[prb,aps,reprint,showpacs,groupedaddress,superscriptaddress,twocolumn,final,prbib,longbibliography]{revtex4-1}

\usepackage{color}
\usepackage{bm}
\usepackage{graphicx}
\usepackage{setspace}
\usepackage{appendix}
\usepackage{epstopdf}
\usepackage{amsmath}
\usepackage{amssymb}
\usepackage{verbatim}
\usepackage{bm}
\usepackage{hyperref}
\hypersetup{
colorlinks=true,
urlcolor= blue,
citecolor=blue,
linkcolor= blue,
bookmarks=true,
bookmarksopen=false,
pagebackref=true}

\newcommand{\micro}[1]{$\mathrm{\mu}${#1}}
\newcommand{\um}{\micro{m}}
\newcommand{\nm}{\mathrm{nm}}

\newcommand{\Ba}{$\mathrm{BaFe_2As_2}$}
\newcommand{\CoBa}{$\mathrm{Ba(Fe_{1-x}Co_x)_2As_2}$}
\newcommand{\PBa}{$\mathrm{BaFe_2(As_{1-x}P_x)_2}$}
\newcommand{\KBa}{$\mathrm{Ba_{1-x}K_xFe_2As_2}$}

\newcommand{\FeSe}{$\mathrm{FeSe}$}
\newcommand{\tc}{T_C}
\newcommand{\Tc}{$\tc$}
\newcommand{\App}[1]{Appendix~\ref{#1}}
\newcommand{\fig}{Fig.~}

\newcommand{\Fig}[1]{\fig\ref{#1}}

\newcommand{\Eq}[1]{Eq.~\ref{#1}}

\newcommand{\KK}{Kogan and Kirtley \cite{Kogan2011}}
\newcommand{\etal}{\textit{et al.}}
\newcommand*{\citen}[1]{%
	\begingroup
	\romannumeral-`\x 
	\setcitestyle{numbers}%
	\cite{#1}%
	\endgroup   
}
\begin{document}

\title{Diamagnetic Vortex Barrier Stripes in Underdoped $\mathrm{\mathbf{{BaFe_2(As_{1-x}P_x)_2}}}$}

\author{A.~Yagil}
\affiliation{Department of Physics, Technion - Israel Institute of Technology, Haifa, 32000, Israel}
\author{Y.~Lamhot}
\affiliation{Department of Physics, Technion - Israel Institute of Technology, Haifa, 32000, Israel}
\author{A.~Almoalem}
\affiliation{Department of Physics, Technion - Israel Institute of Technology, Haifa, 32000, Israel}
\author{S.~Kasahara}
\affiliation{Department of Physics, Kyoto University, Kyoto 606-8502, Japan}
\author{T.~Watashige}
\affiliation{Department of Physics, Kyoto University, Kyoto 606-8502, Japan}
\author{T.~Shibauchi}
\affiliation{Department of Advanced Materials Science, University of Tokyo 5-1-5 Kashiwanoha, Kashiwa, Chiba 277-8561, Japan}
\author{Y.~Matsuda}
\affiliation{Department of Physics, Kyoto University, Kyoto 606-8502, Japan}
\author{O.~M.~Auslaender}\email[]{ophir@physics.technion.ac.il}
\affiliation{Department of Physics, Technion - Israel Institute of Technology, Haifa, 32000, Israel}

\begin{abstract} 
We report magnetic force microscopy (MFM) measurements on underdoped \PBa\ ($x=0.26$) that show enhanced superconductivity along stripes parallel to twin boundaries. These stripes of enhanced diamagnetic response repel superconducting vortices and act as barriers for them to cross. The width of the stripes is hundreds of nanometers, on the scale of the penetration depth, well within the inherent spatial resolution of MFM and implying that the width is set by the interaction of the superconductor with the MFM's magnetic tip. Unlike similar stripes observed previously by scanning SQUID in the electron doped \CoBa, the stripes in the isovalently doped \PBa\ disappear gradually when we warm the sample towards the superconducting transition temperature. Moreover, we find that the stripes move well below the reported structural transition temperature in \PBa\ and that they can be much denser than in the \CoBa\ study. When we cool in finite magnetic field we find that some vortices appear in the middle of stripes, suggesting that the stripes may have an inner structure, which we cannot resolve. Finally, we use both vortex decoration at higher magnetic field and deliberate vortex dragging by the MFM magnetic tip to obtain bounds on the strength of the interaction between the stripes and vortices. We find that this interaction is strong enough to play a significant role in determining the critical current in underdoped \PBa. 
\end{abstract}

\maketitle

\section{\label{sec:intro}Introduction}
One of the hallmarks of the emergent nematic phase in the iron-based superconductors (Fe-SCs) \cite{Wen2011,Stewart2011,Chubukov2015} are twin boundaries \cite{Hilgenkamp2002} (TBs) that appear in underdoped samples as they are cooled through the transition from the high temperature tetragonal phase to the low temperature orthorhombic phase \cite{Chu2009,Tanatar2009,Prozorov2009,Chu2010,Kasahara2012,Chu2012,Bohmer2012,Tanatar2013}. TBs are not limited to Fe-SCs -- they occur in many superconductors, including cuprates \cite{Dolan1989,Vinnikov1990,Roitburd1990,Gammel1992,Duran1992,VlaskoVlasov1994,Duran1995,Welp1995,MaggioAprile97,Herbsommer2000,Figueras2006,Palau2006,Shapira2015}, and are important for several reasons. In Fe-SCs their properties encode information about the nature of the superconducting phase and its competing orders \cite{Chuang2010,Rosenthal2014,Watashige2015}. On a more practical level, TBs play a crucial role in the way superconducting vortices move through a superconductor. The dissipative motion of vortices,  quantized whirlpools of charge encircling a core with suppressed superconductivity, is a limiting factor in applications of type-II superconductors \cite{Larbalestier01,Scanlan04,Foltyn2007}. Thus, understanding how TBs affect vortices is important for developing superconductor technologies.

\begin{figure*}[btp]
\centering\includegraphics[width=1.75\columnwidth]{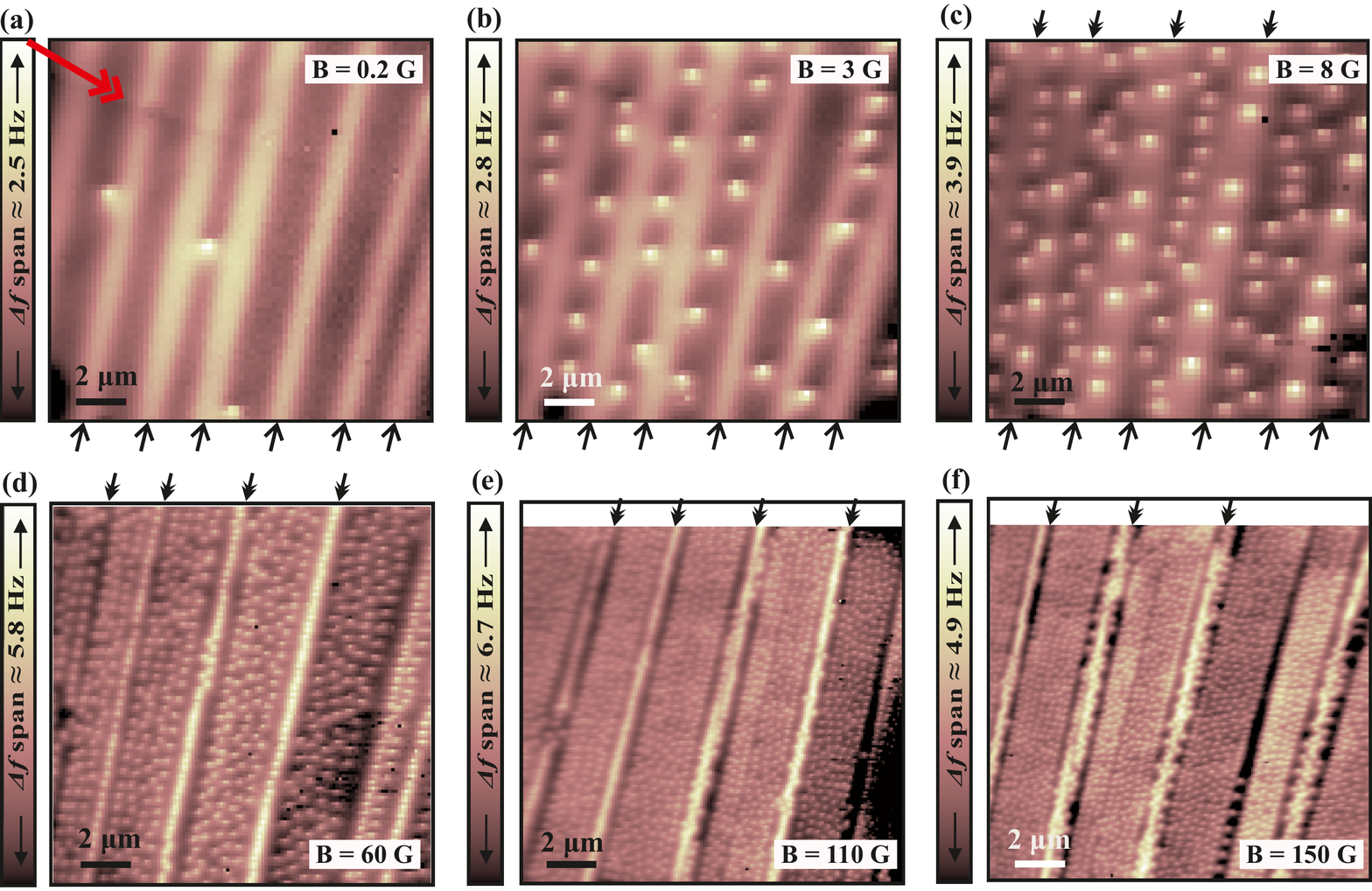}
\caption{Images showing the relationship between vortices and stripes at $T=4.5$~K. The magnetic field is indicated on each panel. In all panels we subtracted a plane from the data.
\textbf{(a)} Image showing three vortices and bright stripes (marked by arrows on the bottom) on a dark background. We use a scratch (marked by a double-headed arrow in the upper left corner) to align images to each other.  
\textbf{(b)} Image showing that vortices avoid the bright stripes. The arrows on the bottom are copied from (a) and show that the stripes have not moved. %
\textbf{(c)} After field-cooling in a higher field some vortices appear in the middle of the bright stripes. The stripes here (marked by arrows on top) are not the same stripes that appeared in (b) (bottom arrows).  %
\textbf{(d)} More vortices accumulate both on and off the stripes when we field-cool in a higher field. The arrows were copied from panel (c) to show that the stripes have not moved relative to one another. 
\textbf{(e)} The stripes are the same as in (c), (d), as shown by the arrows that were copied from panel (c). 
\textbf{(f)} This scan area is slightly shifted with respect to (a-e) but some stripes are the same as in (c-e), as highlighted by the arrows from panel (c). 
[The scan heights, which do not qualitatively affect the images, are $130$~nm, $170$~nm, $100$~nm, $100$~nm, $70$~nm, and $110$~nm, respectively for panels (a-f).]}
\label{fig:vortex_decoration}
\end{figure*}

Frequently the superfluid density ($\rho_s$) is suppressed on a TB \cite{Blatter94}. When this happens the TB acts as a pinning site because of the reduced energetic cost of locating a vortex core on it. Such pinning behavior has been observed in the cuprates \cite{Dolan1989, Vinnikov1990, Gammel1992, MaggioAprile97, Shapira2015} where TBs also act as channels that are easy for vortices to move along, and hard for them to cross \cite{Duran1992, VlaskoVlasov1994, Duran1995, Welp1995, MaggioAprile97, Shapira2015}. 
Other behavior is also possible. For example, in both conventional low-\Tc\ (superconducting transition temperature) superconductors \cite{Khlyustikov1987}, as well as in cuprates \cite{Inderhees1988,Abrikosov1989a,Abrikosov1989b}, there have been reports of enhanced \Tc\ near TBs. This implies that vortices can be repelled from TBs. 

In the Fe-SCs the impact of TBs on superconductivity is different in different materials. For example, scanning tunneling microscopy (STM) experiments on TBs in \FeSe\ have reported a reduced gap as well as vortex pinning in both thin films\cite{Song2012} and in single crystals\cite{Watashige2015}. Kalisky \etal\cite{Kalisky2010} used a scanning superconducting quantum interference device (SQUID) to show stripes of enhanced diamagnetic response in underdoped but not in overdoped \CoBa. Kirtley \etal\cite{Kirtley2010,Kogan2011} showed that these results are consistent with an enhancement of $\rho_s$ on thin sheets embedded in the bulk sample. Finally, also using SQUID microscopy, Kalisky \etal \cite{Kalisky2011} showed that vortices tend to avoid the stripes of enhanced superconducting response and that, when manipulated, they tend to move parallel to the stripes rather than to cross them. However, the scanning SQUID results were resolution-limited to $\approx2$~\um, much larger than the in-plane penetration depth $\lambda_{ab}$, which is a few hundred nanometers in \CoBa\cite{Luan2011}. Unexpectedly, high-resolution magnetic force microscopy (MFM) on nominally identical samples did not detect similar stripes \cite{Luan10}.

We chose to study \PBa\ for its outstanding properties. First, \PBa\ is less disordered \cite{Shishido2010, Klintberg2010, Demirdis2013} than other members of the \Ba\ family. This is due to the doping being isovalent -- unlike electron doped \CoBa\ and hole doped \KBa, the charge density in \PBa\ does not change with $x$. A second notable property is unconventional behavior near $x_\mathrm{opt}$ \cite{Shibauchi2014}, including a peak in $\lambda_{ab}(x)$  \cite{Hashimoto2012,Lamhot2015}. The origin of this highly unusual and surprising effect is still under debate\cite{Levchenko2013, Fernandes2013, Chowdhury2013, Nomoto2013, Chowdhury2015}. In other respects \PBa\ is a typical member of the \Ba\ family. The parent compound is a metal that undergoes magnetic and structural phase transitions at $T_N=T_S\approx135$~K. Upon doping $T_N$ and $T_S$ decrease and diverge until they drop sharply near $x\approx0.3$. 
The superconducting $\tc(x)$ is domed, rising from zero at $x\approx0.2$ to maximum at $x_\mathrm{opt}\approx0.3$ and dropping to zero again at $x\approx0.7$\cite{Hashimoto2012}.

Here we present MFM measurements on underdoped \PBa. We find features with an enhanced diamagnetic response running parallel to TBs. These show up as stripes of enhanced diamagnetic response at low magnetic field. The stripes disappear as the sample is warmed towards \Tc. When the sample is cooled in a finite magnetic field, vortices favor the regions off the stripes. When we use the magnetic tip of the MFM to deliberately try to drag individual vortices across the stripes, they act as barriers. Finally -- we find that the stripes are mobile even below \Tc, although it is below the reported $T_S$ for our underdoped sample.

Much of the phenomenology of our observations agrees with the SQUID results of Kalisky, Kirtley \etal \cite{Kalisky2010,Kirtley2010,Kalisky2011} on \CoBa.
This is significant for several reasons. First, we provide confirmation of  stripes of enhanced diamagnetic response in a material other than \CoBa. Second, the higher spatial resolution of our measurements allows us to show that the scale of both the modulated diamagnetic response as well as the vortex repulsion is given by $\lambda_{ab}$. Overall, and despite the different spatial scale for the stripes that we find in \PBa, our observations validate the interpretation and analysis put forth by Kalisky, Kirtley \etal \cite{Kalisky2010,Kalisky2011}. We also find important differences between the stripes in the two materials. The most important of these is that in \PBa\ the stripes decay when we increase temperature whereas in \CoBa\ they are enhanced.

\section{\label{sec:experiment}Experiment}

\subsection{Sample}
Our sample is a high-quality single crystal with an area of $\approx0.25\mathrm{mm}^2$ and thickness tens of microns, grown by the self-flux method and annealed in vacuum \cite{Kasahara2010}. This sample is part of a series spanning the superconducting dome that we reported on previously as part of a study of the dependence of $\lambda_{ab}$ on doping \cite{Lamhot2015}. The sample was cleaved and analyzed by energy-dispersive x-ray spectroscopy (EDS) to determine the doping $x$ at the actual scanned surface at several different locations using a measurement area of $\approx~50~\mu\mathrm{m}~\times~50~\mu\mathrm{m}$. In addition to $x$, the EDS reported the expected atomic compositions for Ba (19.0\%-–21.0\%) and Fe (38.4\%-–41.0\%). The scatter of the values we obtained for $x=0.26$ by EDS gives a variance of $\delta x<0.01$.

In this work we concentrate on a moderately underdoped sample with $x=0.26$. At this doping we determined that $\lambda_{ab}=220\pm20$~nm and $\tc\approx22$~K \cite{Lamhot2015}. The main source of error in $\lambda_{ab}$ and in \Tc\ at $x=0.26$ is our method of measurement -- the results were the same in multiple regions, indicating that the sample is very uniform. In this sample, as well as in another $x=0.26$ sample, we observed vortices lining up at roughly $45^\circ$ to the crystal a-b axes as indicated by room temperature electron backscatter dispersive spectroscopy (EBSD). The measured orientation of the stripes indicates that they are parallel to TBs. In this work we study those stripes further.

\subsection{Measurement}
\begin{figure}
\centering\includegraphics[width=1\columnwidth]{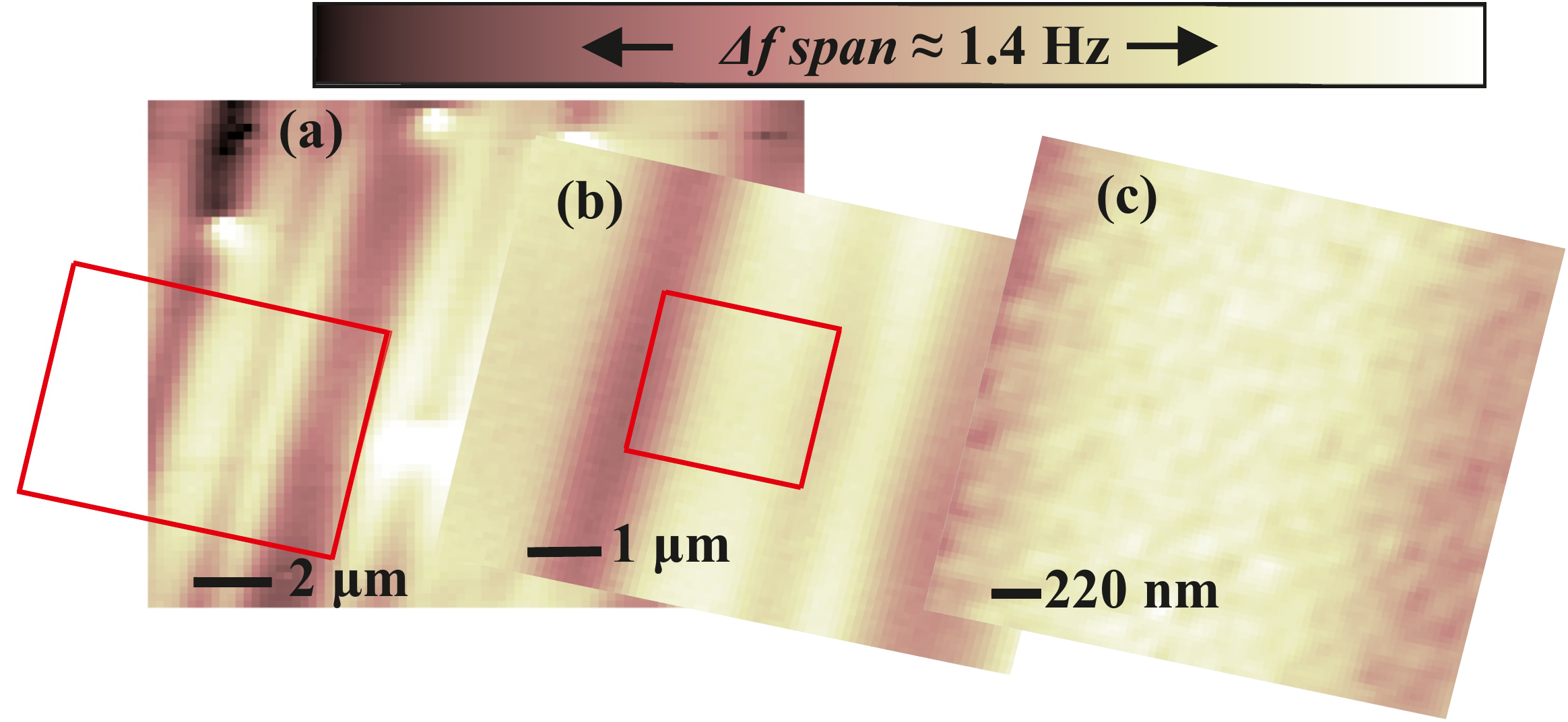}%
\caption{A series of scans zooming in on a stripe at $T=4.5$~K and $0<B<0.5$~G. In (a,b) a plane has been subtracted from the data. \textbf{(a)} A large-range scan with $h=200$~nm. The bright disks are vortices, and the stripes are visible as bright lines. The rectangle frames the zoom-in area shown in (b). \textbf{(b)} A scan with $h=100$~nm. The scan area is the red rectangle in (a). Note the two stripes, where one is wider than the other. The rectangle shows the zoom-in area shown in (c). \textbf{(c)} A scan with $h=70$~nm. The scan area is the red rectangle in (b). The length of the scale-bar is $220~\nm\approx\lambda_{ab}$, which gives the scale for the limit of our resolution for magnetic imaging. The small corrugations are topographic features. If the stripes have internal structure, it is below our resolution or sensitivity.}
\label{fig:stripe_zoom}
\end{figure}
We use frequency modulated MFM to measure the resonance frequency ($f$) of a cantilever holding a magnetic tip that we scan across a sample. The resulting map of the spatial dependence of the frequency shift $\Delta f=f-f_0$ gives a map of a derivative of the vertical force acting on the tip:
 \begin{equation*}
 \Delta f\approx -\frac{f_0}{2k}\frac{\partial F_z}{\partial z},
 \end{equation*}
where $f_0$ is the resonance frequency in the absence of a sample and $k$ is the cantilever spring constant \cite{note:Tip}. The components of the force  $\vec{F}=\vec{F}_{lat}+F_z\hat{z}$ are not directly imaged but can be estimated from the MFM signal by assuming a model for the tip\cite{Ophir09}.

After we set the tip-sample voltage ($V_{t-s}$) to compensate for the contact potential difference ($V_{t-s}^*$), $\vec{F}$ is predominantly magnetic and comes from two main sources. The first is the superconducting screening currents that are responsible for the Meissner shielding of the magnetic field exerted by the tip. The second is the interaction between the magnetic tip and magnetic field from superconducting vortices. We make use of both forces in this work. The first gives information on the strength of superconductivity, as encoded in $\lambda_{ab}$. The second gives information on material defects which can attract or repel vortices.

Most of the results we report below come from two kinds of experiments: imaging and vortex manipulation. For both we cool the sample through \Tc\ at finite magnetic field (field-cool\cite{note:fieldcool}) along the crystal c-axis. For this we retract the magnetic MFM tip from the surface to make sure it does not influence the sample during the field-cool cycle. Once the sample has stabilized at the desired $T<\tc$ we proceed to scan the tip at a constant height ($h$) above the surface.

We scan in one of three modes. In the first we set $V_{t-s}$ to several volts. This is much larger than $V_{t-s}^*$ and allows us to be very sensitive to surface topography variations. The other two modes are magnetic. For these we set $V_{t-s}=V_{t-s}^*$ to significantly enhance our sensitivity to the magnetic interaction between the tip and the sample at the expense of the electrostatic interaction.

The magnetic modes of operation are surveillance and manipulation \cite{Straver08}. In the latter we bring the tip so close to superconducting vortices that the force the tip exerts can move them away from the pinning site they are trapped in. This is useful because the way vortices move can reveal information that is not available to other surface sensitive techniques\cite{Straver08,Ophir09,Zhang2015,Shapira2015}.

In surveillance mode we retract the tip far enough from the surface for the forces the tip exerts to be too weak to depin vortices in the sample. We then scan the tip in a raster pattern on a plane parallel to the surface and obtain a magnetic image that maps the vortex locations and measures the local diamagnetic interaction between the tip and the sample \cite{Luan10,Luan2011,Lamhot2015,Zhang2015}. This gives local information on the absolute value of $\lambda_{ab}$ and through it on $\rho_s\propto\lambda_{ab}^{-2}$.

\section{\label{sec:results}Results}
\begin{figure}
\centering\includegraphics[width=1\columnwidth]{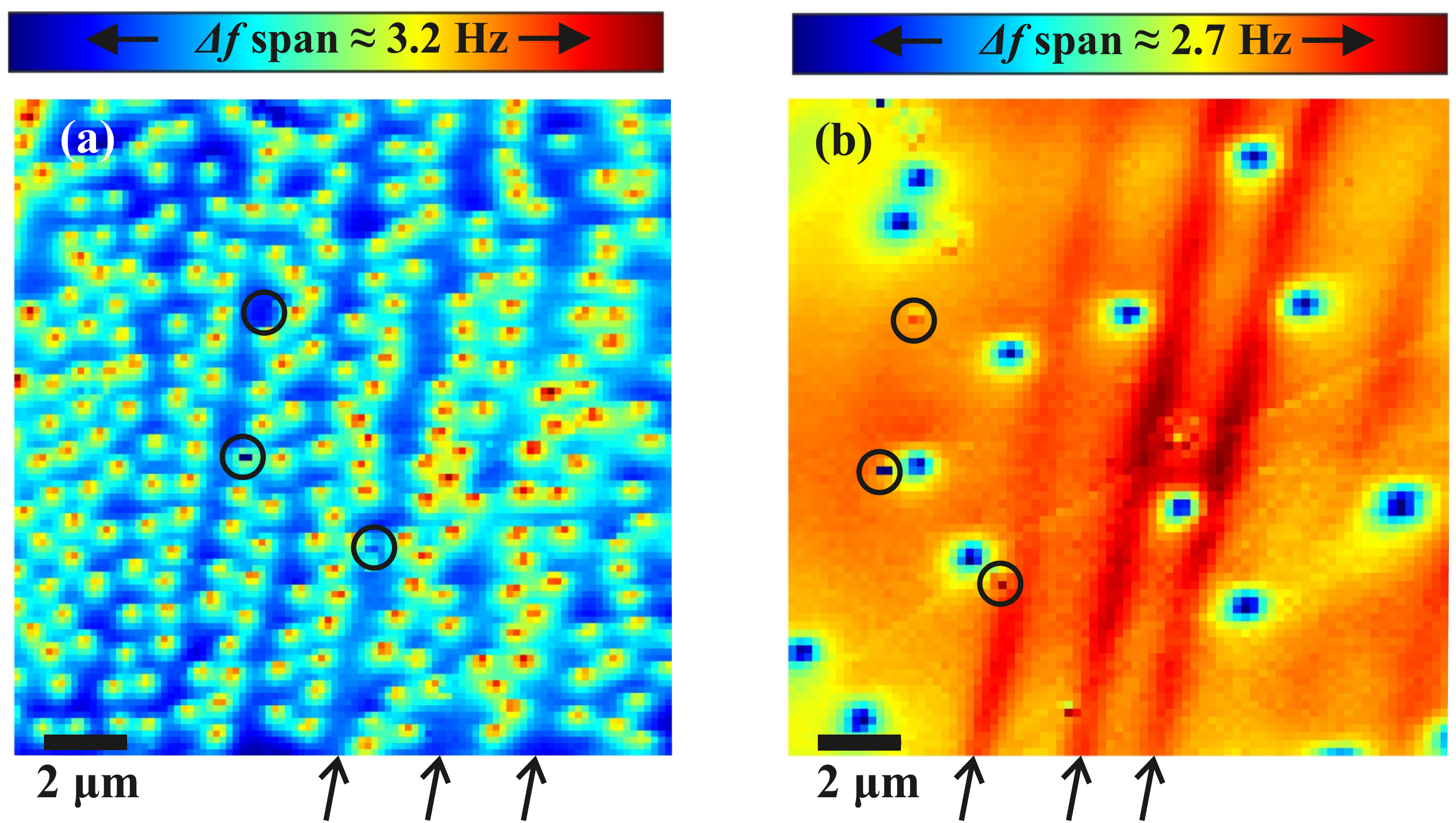}%
\caption{Comparison between cooling in positive (a) and negative (b) field. The circles mark surface features we use to align the scans to each other. In both scans, which were taken at $T=4.5$~K, a plane has been subtracted from the data. %
\textbf{(a)} $B=20$~G image. The arrows mark stripes that are easy to identify in (b). Clearly repulsive vortices avoid the stripes. %
\textbf{(b)} $B=-1.3$~G image of the same area as in (a). Visible are attractive vortices that clearly avoid the stripes. The arrows mark the stripes that are also visible in (a).  [The scan heights, which do not qualitatively affect the images, are 60 nm, 115 nm, respectively for panels (a,b).]}
\label{fig:vortex_defects}
\end{figure}

\subsection{Imaging}
Many of our results can be seen in \Fig{fig:vortex_decoration} which shows the same area in the sample after we field-cool at different values of magnetic field. One can clearly see a modulation of the MFM signal along stripes that we have previously determined to be parallel to TBs in this material \cite{Lamhot2015}. This modulation (and an occasional similar modulation rotated by $90^\circ$, see e.g. \Fig{fig:90deg} in \App{app:stripes90}) appears only in parts of the sample. Figure~\ref{fig:vortex_decoration} also features vortices that are repelled from the MFM tip and appear as regular disks with a signal more positive than the background.

The field progression in \Fig{fig:vortex_decoration} shows several effects. The first are the bright stripes, which exist even at low field [\Fig{fig:vortex_decoration}(a)], that correspond to an enhanced diamagnetic response of the superconductor. The stripes can have a half-width down to a scale of $\lambda_{ab}$ but this can vary from stripe to stripe, as can be seen in \Fig{fig:stripe_zoom}(b). When we cool the sample in a slightly higher field vortices freeze preferentially between the bright stripes [\Fig{fig:vortex_decoration}(b)], indicating that the stripes are energetically unfavorable for vortices. The vortices off the bright stripes appear to form lines but this is due to the distance between the bright stripes which in this particular region happens to be on the scale of several $\lambda_{ab}$.

When we cool in an even higher field some vortices nucleate in the middle of the wider bright stripes [\Fig{fig:vortex_decoration}(c)]. Since the scale for vortex pinning is set by the coherence length $\xi_{ab}$, which is in the nanometer range \cite{Song2012}, this suggests  that the stripes have internal structure -- what we see as single stripes may actually be several stripes too close for us to resolve. This picture is supported by the apparent straight lines along which the vortices seem to be organized (if the gap between stripes were wider we would expect a line connecting the vortices to meander more). To study this point further we zoomed in on areas with stripes. We show an example in \Fig{fig:stripe_zoom}, where the stripes are not identical. When we zoom in on a particularly wide stripe we still cannot resolve any internal structure. This means that it is either absent or on a scale that is not accessible to us, because the resolution of MFM imaging of the superconducting response is limited by $\lambda_{ab}$.

The rest of the panels in \Fig{fig:vortex_decoration} show what happens in yet higher field: vortices fill the areas between the stripes and coalesce in high density along lines. Presumably these high vortex density lines are the result of vortices being repelled from adjacent stripes, which are also apparent. Without the low-field scans it would be very easy to conclude from such scans that vortices prefer the areas between the wider parts. This has also been the interpretation of vortex decoration data in \KBa\cite{Yang2012} and in \PBa\cite{Vinnikov2014} as well as magnetization data in \CoBa \cite{Prozorov2009}.  Only by looking at very low field can we conclude that vortices avoid the bright stripes.

\begin{figure*}
\centering\includegraphics[width=2\columnwidth]{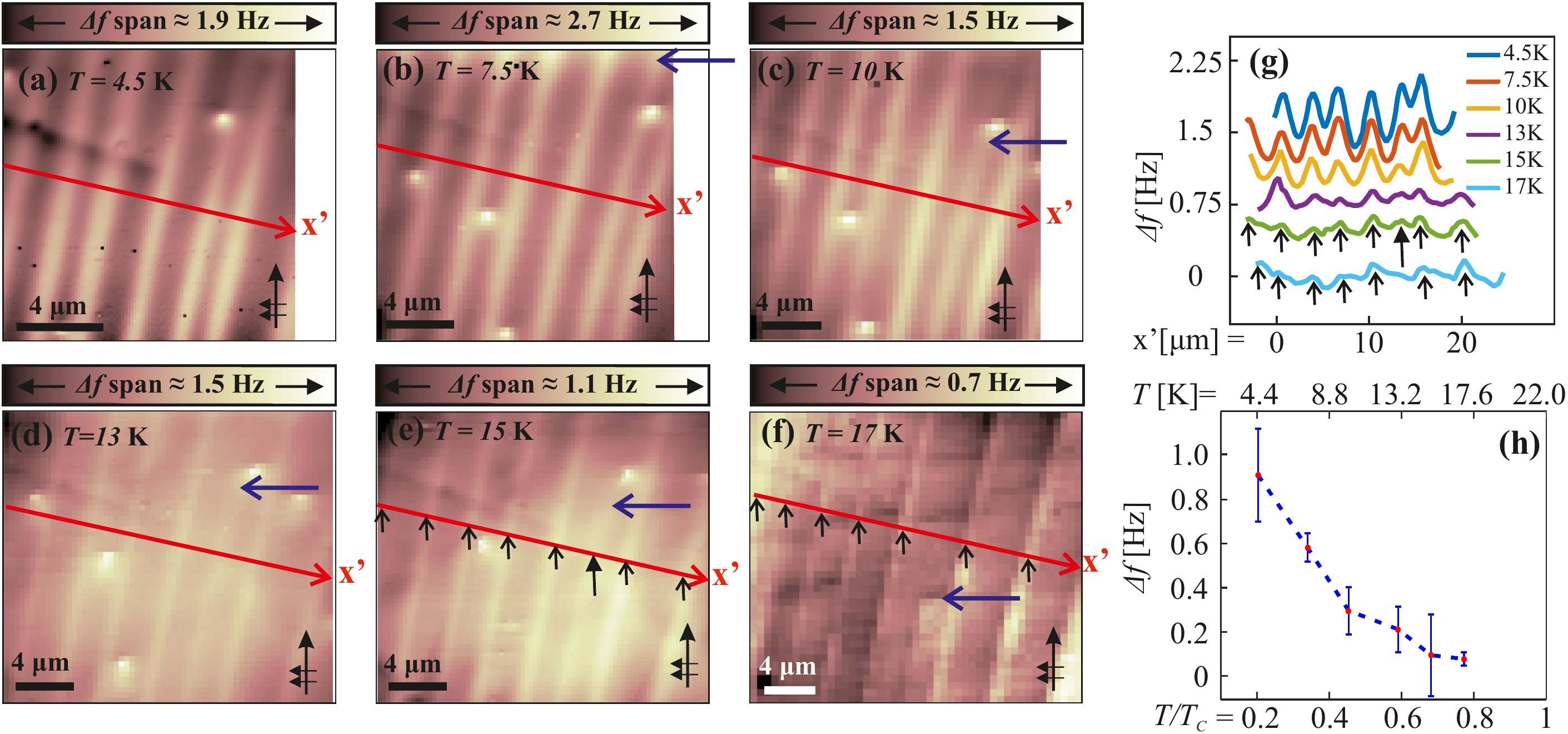}
\caption{Strength of the stripes as a function of temperature for one area. Panels (a-f) show maps of $\Delta f$ at increasing temperatures. A plane fit has been subtracted from each image. The red lines show the trajectories of the traces shown in (g). We changed the field\cite{note:fieldcool} after acquiring the scan in (a) but kept it constant for (b-f). In panels (a-f) the double left-pointing arrows show the fast scan direction and the long upward pointing arrow shows the slow scan direction along which the tip was incremented after each fast scan period of the raster pattern. The blue horizontal arrow points out a stripe that moved at $T<\tc$.  \textbf{(a)} Scan for $T=4.5$~K, \textbf{(b)}  $T=7.5$~K, \textbf{(c)} $T=10$~K, \textbf{(d)} $T=13$~K, \textbf{(e)} $T=15$~K, and \textbf{(f)} $T=17$~K. %
[The scan heights for (a-f), which do not qualitatively affect the images, are $220$~nm, $180$~nm, $180$~nm, $180$~nm, $190$~nm, and $210$~nm, respectively for panels (a-f).] %
\textbf{(g)} The signal along the red lines in scans (a-f) after subtracting a parabolic background and aligning based on surface features. For clarity the curves are offset from each other by $1.7,~1.4,~1.1,~0.8,~0.5,~0$~Hz. \textbf{(h)} The average amplitude (peak-to-peak) of the stripes for scans with $h\approx150$~nm versus $T/\tc$ and $T$. The error bars give our estimate for 70\% confidence intervals\cite{note:p2p}. The line is a guide to the eye.}
\label{fig:Tdep}
\end{figure*}

Careful inspection of \Fig{fig:vortex_decoration} shows that the stripes can move. While the stripes in (a) appear at the same locations as the stripes in (b) and the stripes in  (c-f) are all also at the same positions, the stripes in (b) are not the same stripes we see in (c). We are sure this is not due to an offset of the scan area because we see the same topographic features in both scans. As explained in Sec.~\ref{sec:experiment} and in [\citen{note:fieldcool}], for each field-cool we heated the sample to $T>\tc$ prior to applying the new field. We chose $T=25$~K, lower than the reported\cite{Kasahara2012,Bohmer2012} $T_S\approx45$~K for samples with $x=0.26$. As long as $T\ll T_S$ the domains should not be affected by heating but $T_S(x)$ follows a very steep curve near $x=0.26$ so the $\approx45$~K has a large error bar and may be as low as $\approx35$~K \cite{Bohmer2012}. Consequently, at $T=25$~K the domains may not be totally frozen in. We thus speculate that the shift of the stripes between (b) and (c) is due to heating of the sample during a field change.

The morphology of the stripes and their impact on vortices do not depend on the polarity of the applied field along the c-axis. For example, \Fig{fig:topo}(a) in \App{app:topo} shows the same stripes that appear in \Fig{fig:vortex_decoration}(a) even though the field is opposite, as evidenced by the presence of vortices that are attracted to the tip (attractive vortices), which are the vortices we obtain when we cool in a negative field. That the interaction between vortices and the stripes does not depend on the field orientation, can also be seen in \Fig{fig:vortex_defects}, which shows that attractive vortices avoid the bright stripes just like vortices that are repelled from the tip (repulsive vortices).

Figure~\ref{fig:Tdep} shows that the contrast of the stripes decays with increasing $T$, until we lose track of them just below \Tc. In \Fig{fig:Tdep}(a-f) we show repeated scans of the same area with increasing $T$ up to $17$~K, which show the stripes very clearly. In scans at $T=20$~K taken under comparable conditions we did not detect any stripes. For the images in \Fig{fig:Tdep}(a-f) we changed the field after taking the scan in (a) but once it was set for the scan in (b) we did not change it. This means that the sample spent time at $T>\tc$ only between the scan in (a) and the rest of the scans\cite{note:fieldcool}. Conveniently the scan area contains the same scratch from \Fig{fig:vortex_decoration}. This scratch allows us to align the scans to one another so that we can extract the MFM signal along the same line [red line in \Fig{fig:Tdep}(a-f)], that was chosen to be away from the scratch, as well as vortices.  In \Fig{fig:Tdep}(g,h) we quantify the decay of the amplitude of the stripes with $T$. Figure \ref{fig:Tdep}(g) shows the signal from the stripes along the red lines in (a-f). In \Fig{fig:Tdep}(h) we combine information from several scans at each temperature to plot the average peak-to-peak amplitude of the stripes. We use several scans for each $T$ in order to compare data at the same height for all temperatures \cite{note:p2p}. The error bars in \Fig{fig:Tdep}(h) give our estimate for 70\% confidence intervals and reflect both the variation from stripe to stripe and errors we introduce in the data analysis.

Figure~\ref{fig:Tdep} shows evidence for stripes moving at $T<\tc$. The motion occurs in the stripe marked by the large, left pointing, blue arrow. In panels (a,b) the stripe extends across the whole scan area but in (c) one can see that the stripe terminates in the middle of the image (at the arrow). In panels (d-e) the stripe remains stationary but in panel (f) we see that it has retracted even more towards the bottom of the scan area, as indicated by the position of the arrow. Similar partial stripes appear in Fig.~3 in Kalisky, Kirtley \etal\ \cite{Kalisky2010}. We speculate that such terminating stripes are needle-shaped domains, which means that they are composed of two TBs running parallel to each other and another short TB running across at the termination point. The analysis we present in Sec.~\ref{sec:discussion} and in \App{app:KK} supports this claim. If the terminating stripes are domains bounded by TBs then the fact that they can move at $T<\tc$ is an indication that TBs are weakly pinned in our sample.

\subsection{Vortex manipulation}\label{Sec:discussion}
\begin{figure}
\centering\includegraphics[width=1\columnwidth]{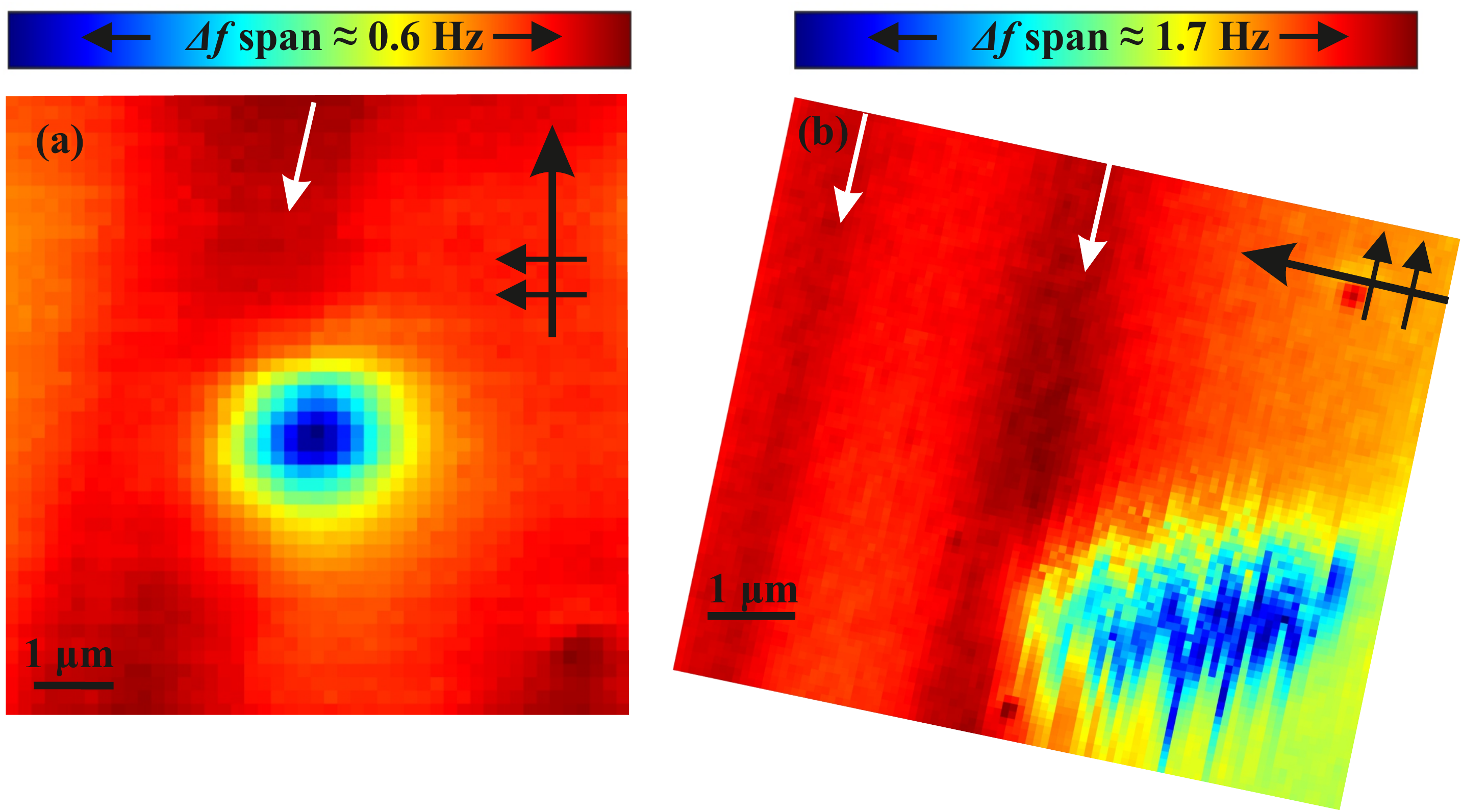}
\caption{Experiment which shows that it is hard to drag vortices across the stripes (which are marked by white downward arrows). The crossed arrows show the raster pattern for this scan (in each scan the thin double arrows show the fast scan direction and the thick arrow shows the slow scan direction). For both scans $T=12$~K, $B=-0.3$~G (attractive vortices). \textbf{(a)} A surveillance scan ($h=600$~nm). At this scan height the force applied by the tip is smaller than the vortex pinning force, as can be seen from the regular shape of the vortex.  For this plot we subtracted a plane from the raw data. %
\textbf{(b)} Manipulation scan for the same vortex ($h=150$~nm). The elongated shape of the vortex in the direction of the slow scan axis indicates that the vortex moves as a result of the force exerted by the tip. Even though the tip exerted sufficient force in order to move the vortex we were not able to drag it across the stripe. Note that in this scan two stripes can be seen.}
\label{fig:dragging}
\end{figure}
Close inspection of \Fig{fig:Tdep}(a-f) shows that the vortex configuration changes from scan to scan and that some of the vortices move mid-scan. Similar effects are common in vortex MFM \cite{Moser98} and are frequently considered a disadvantage of this technique. Here we used this ability deliberately \cite{Straver08,Ophir09,Luan10,Shapira2015} to attempt to drag vortices across stripes. To this end we cooled the sample in a field oriented to give attractive vortices. After mapping out their locations with respect to the stripes we heated the sample to reduce vortex pinning, brought the tip closer to the surface and proceeded to scan at values of $h$ where we saw significant vortex motion. Figure~\ref{fig:dragging}  shows one example of an attempt to drag vortices across the stripes. %
We estimate that in this scan the scale for the maximum lateral force we exerted on the vortex was $F_\mathrm{lat}^\mathrm{max}\approx10$~pN. For this estimate we use the typical tip parameters that are listed below \Eq{eq:dFzcone} in \App{app:KK}. %
We performed such dragging attempts many times and at different temperatures. In all of these cases it was clear that the stripes act as barriers for vortices, in agreement with the results of Kalisky, Kirtley \etal\ on \CoBa\cite{Kalisky2011}.

\section{\label{sec:discussion}Discussion}
Despite the difference in scale, The stripes we report are similar to those reported by Kalisky, Kirtley \etal\cite{Kalisky2010,Kalisky2011} in \CoBa. Following similar reasoning we conclude that the enhanced diamagnetic response we observe is therefore probably also due to TBs. 
Our results validate the interpretation of the results on \CoBa\cite{Kirtley2010,Kogan2011} which did not explain why contemporaneous MFM measurements did not detect such stripes\cite{Luan10}. We speculate that TBs did not appear in MFM scans of \CoBa\ because they do not appear everywhere in the sample. Even in this work we saw stripes only in some areas. This is probably due to inhomogeneous strain induced in the sample by thermal contraction.

As a test for our interpretation we extend analysis put forth by \KK\ for SQUID microscopy to MFM. The derivation is in \App{app:KK} where we also apply the analysis to our data. As shown in \Fig{fig:simulation} in \App{app:KK}, the \KK\ model is consistent with our data and the comparison gives model parameters similar to the \CoBa\ model parameters.

There are also differences between our results and the results on \CoBa\cite{Kalisky2010,Kirtley2010,Kalisky2011}. The first of these is that our measurements indicate that stripes can be composed of several TBs. The evidence for this is threefold. First, our stripes are not all the same (see e.g. \Fig{fig:stripe_zoom}) -- some are wider and some are brighter. Second, when we field-cool vortices nucleate in the middle of some of the stripes but not in others [e.g. \fig\ref{fig:vortex_decoration}(c)], indicating that different stripes can have different internal structure. Third, there are stripes that terminate in the middle of a scan area (\fig\ref{fig:Tdep}). This indicates that at least some of the stripes correspond to narrow domains. This interpretation is further supported by the analysis in \App{app:KK}.

Another important difference between our results and the results on \CoBa\ is the temperature dependence\cite{Kalisky2010,Kirtley2010}. In \CoBa\ the stripes were enhanced when the sample was heated, although they were absent above $\tc$. A fit of the temperature dependence suggested that \Tc\ on the \CoBa\ stripes was higher than the bulk \Tc. Here we see a very clear decay of the stripes [\fig\ref{fig:Tdep}(g,h)], which disappear at $T<\tc$.

The contrasting observations on TBs can be considered in the context of other differences between \CoBa\ and \PBa. In \CoBa\ $\lambda_{ab}^{-2}$ increases monotonically from the underdoped edge of the superconducting dome to $x_\mathrm{opt}$ \cite{Luan2011}. The enhanced superconducting response on stripes might lead one to hypothesize that in \CoBa\ TBs attract electrons and make the effective doping on a TB closer to optimal than the bulk underdoped sample \cite{Li2013}. In \PBa\ the mechanism has to be different because $\lambda_{ab}^{-2}$ has a minimum near $x_\mathrm{opt}$\cite{Hashimoto2012,Lamhot2015}.

We can rule out several alternative explanations for the stripes. The lack of dependence of the properties of the stripes on the direction of the magnetic field along the c-axis implies that the stripes are likely not signatures of magnetic domains, which would have flipped with the field after a field-cool\cite{note:fieldcool}. This independence also rules out the stripes as the signature of Josephson junctions between domains, that have been invoked to explain the properties of TBs in cuprates \cite{Deutscher1987}.

We can also place an upper bound on topographic variations associated with TBs. As explained in \App{app:topo}, if there are height variations associated with stripes they are well below our nanometer-scale resolution for height variations. This is in-line with the $10$~pm scale for the height variations measured across TBs by STM in \FeSe\ thin films \cite{Song2012}. We can also rule out the accumulation of localized charge on the TBs. Since our tip is metallic this would give rise to attraction rather then the repulsion we detect.

Finally, we note that the impact of TBs on superconductivity depends on material parameters. For example, it has been shown that the interaction between a vortex and a TB can be either repulsive or attractive \cite{Khlyustikov1987,Abrikosov1989a,Abrikosov1989b}. When TBs have an adverse effect on superconductivity they can act as traps for vortices. This is the case in \FeSe, where the superconducting gap has been shown to be reduced on TBs \cite{Song2012,Watashige2015}. An important difference between \FeSe\ and the $\mathrm{BaFe_2As_2}$ family is that in the latter magnetic order competes with superconductivity whereas in \FeSe\ it does not\cite{Medvedev2009,Terashima2015}. This raises the possibility that the competition with the magnetic phase plays a role in enhancing superconductivity on TBs in the $\mathrm{BaFe_2As_2}$ family. 

\begin{figure}
\centering\includegraphics[width=1\columnwidth]{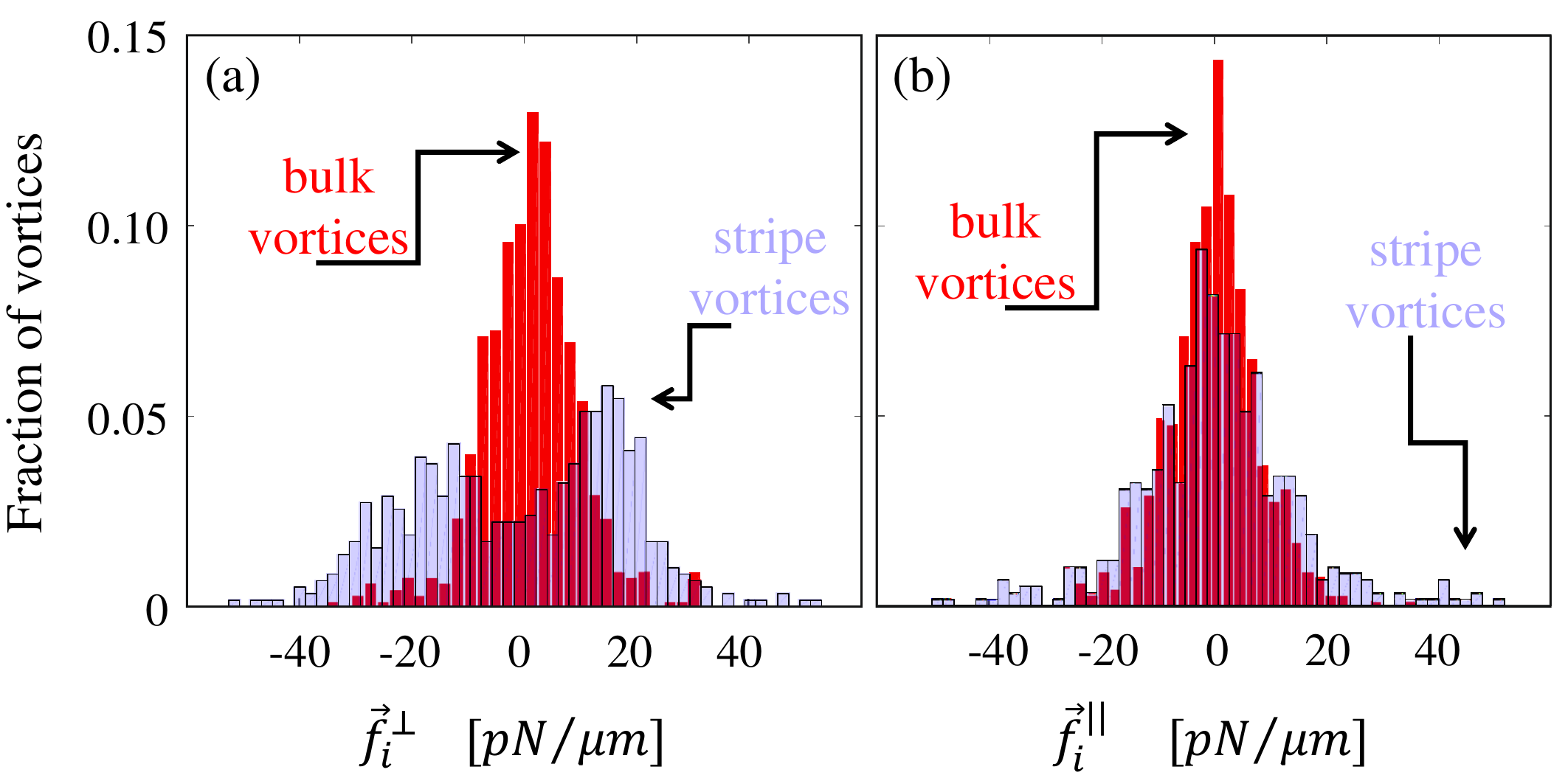}
\caption{Distribution of the vortex-vortex interaction net force per unit length derived by \Eq{eq:Fvv} from the positions of $586$ bulk-vortices and $646$ stripe-vortices in \Fig{fig:vortex_decoration}(f). In the summation in \Eq{eq:Fvv} we included only vortices within a radius of $1.44$~\micro{m}. \textbf{(a)} Histograms of $\vec{f}_i^\perp$. For stripe-vortices the sign gives which side of the stripe they are -- the interaction is always repulsive. \textbf{(b)} Histograms of $\vec{f}_i^\parallel$.}
\label{fig:histograms}
\end{figure}
Regardless of the reason stripes repel vortices in \PBa, we can use vortex decoration to characterize the repulsion. Since the interactions between vortices are well understood\cite{Tinkham75} we can convert vortex positions into the net force per unit length the sample exerts on each vortex in order to keep it stationary ($\vec{f}_i$). To do this we extract the vortex positions $\left\{\vec{r}_i\right\}$ from an image like \Fig{fig:vortex_decoration} and use:
\begin{equation}\label{eq:Fvv}
  \vec{f}_i=\frac{\Phi_0^2}{2\pi\mu_0\lambda_{ab}^3}\sum_{j\neq i} \frac{\vec{r}_i-\vec{r}_j}{\left|\vec{r}_i-\vec{r}_j\right|}K_1\left({\left|\vec{r}_i-\vec{r}_j\right|}/{\lambda_{ab}}\right).
\end{equation}
Here $\mu_0$ is the permeability of free space and $K_1(x)$ is a modified Bessel function of the second kind. In \Eq{eq:Fvv} we ignored the small in-plane anisotropy of the penetration depth. As a result the force between two vortices  depends only on the magnitude of $\lambda_{ab}$.  Figure~\ref{fig:histograms} shows histograms of the results for \Fig{fig:vortex_decoration}(f) where we separate between the component perpendicular to the stripes [$\vec{f}_i^\perp$ in (a)] and along them [$\vec{f}_i^\parallel$ in (b)], as well as between the first row of vortices next to a stripe (stripe vortices, $\vec{f}^s_{i}$) and the rest (bulk vortices, $\vec{f}^b_{i}$). In generating \fig\ref{fig:histograms} we used the low temperature value for $\lambda_{ab}$ and not the value at the unknown vortex freezing temperature. The information we extract is thus for the net force at low temperature, well below the freezing temperature of the vortices.

The effect of the stripes on the force between the vortices is obvious in \Fig{fig:histograms}: on average stripe-vortices experience a repulsion from the stripes and bulk-vortices experience a zero mean, randomly oriented, force. A more quantitative analysis gives information on the pinning force and on the force exerted by the stripes. The net force on each stationary vortex is zero and if we assume the same is true for the force per unit-length then $\vec{f}_{i}+\vec{f}_{p,i}+\vec{f}_{s,i}=0$, where $\vec{f}_{p,i}$ \& $\vec{f}_{s,i}$ are the bulk pinning and stripe repulsion force-densities exerted on vortex $i$. For a bulk-vortex we can assume $\vec{f}_{s,i}\approx0$ so $\vec{f}_{p,i}=-\vec{f}^b_{i}$. As \Fig{fig:histograms} shows, the distribution of $\vec{f}_{p,i}$ is isotropic, has a zero mean, and a standard deviation of $f_p=\langle|\vec{f}^b_{i}|^2\rangle^{1/2}\approx8$~pN/\micro{m}, where $\langle...\rangle$ denotes an average over vortex positions.

The side peaks in \Fig{fig:histograms}(a) show that stripe-vortices experience an average force density of $f^{s,\perp}=|\langle\vec{f}^{s,\perp}_{i}\rangle|=16\pm2$~pN/\micro{m} pushing them away from the stripes [the average was calculated for each of the peaks separately; the error accounts for a 95\% confidence interval as well as for the different positions of the  peaks]. The typical force density that these vortices experience along the stripes is $f^{s,||}=\langle|\vec{f}^{s,||}_{i}|^2\rangle^{1/2}\approx14$~pN/\micro{m}, but like $\vec{f}_{p,i}$, it is randomly directed and averages to zero. The large difference between $f^{s,||}$ and $f_p$ is a reflection of the heavy tails of the $\vec{f}^{s,||}_{i}$ distribution, which are apparent in \Fig{fig:histograms}(b). Note that the values for the force density that we extract from a particular vortex decoration scan give a lower bound on the maximum repulsion a stripe can exert on a vortex. This bound is set by the applied field (in this case $\approx150$~G), which sets the average distance between vortices and hence the scale for vortex-vortex interactions.

It is useful to compare the pinning and stripe repulsion forces to published measurements of the critical current $j_c$. To convert the magnitude of the force density to the magnitude of the current density we use $f=\Phi_0 j$. We can obtain lower bounds for the critical current $j_c$ from the force per unit length we extract from vortex decoration, where there is no vortex motion, or from our vortex dragging attempts. For example, at $T=4.5$~K we obtain from the analysis of \Fig{fig:histograms} that for current flowing along the TBs $j_c\gtrsim0.8~\mathrm{MA/cm^2}$. We obtain a similar scale from the maximum force density we applied in our attempt to drag a vortex across a stripe at $T=12$~K ($f^\mathrm{max}\approx F_\mathrm{lat}^\mathrm{max}/\lambda_{ab}$). This result is similar to the scale in \CoBa\cite{Prozorov2009} and is larger than the value determined in a previous study for underdoped \PBa\cite{Demirdis2013}. 

We are of course not the first to perform vortex decoration in the Fe-SCs. Previous measurements have almost exclusively reported disordered vortex configurations [e.g. in slightly underdoped (MFM\cite{Luan10}) and overdoped (Bitter decoration\cite{Demirdis2011}) \CoBa, in optimally-doped \KBa\ (Bitter decoration\cite{Vinnikov2009}), in optimally-doped and overdoped \PBa\ (Bitter decoration\cite{Demirdis2013,Vinnikov2013,Vinnikov2014}), in slightly underdoped $\mathrm{NdFeAsO_{1-x}F_x}$ (MFM\cite{Zhang2015})]. There have also been reports on vortices organizing along lines in the Fe-SCs. Apart from our MFM work on \PBa\cite{Lamhot2015}, this includes MFM measurements on underdoped \KBa\cite{Yang2012} as well as vortex decoration in optimally-doped \PBa\cite{Vinnikov2014}, where it was speculated that regions of the sample were underdoped. In the last two works the interpretation was that the lines form on the TBs themselves. Here we have shown that, at least in \PBa, this is not the case -- vortices form lines because they are repelled by stripes that are close to one another.

The microscopic origin of the stripes is not clear at this point. Qualitatively they seem to be consistent with numerical results for a two-orbital model \cite{Li2013}. The model gives magnetic domain walls pinned to existing TBs and on which the superfluid density as well as the gap are enhanced. While these results were obtained for electron doping, they are consistent with our results for \PBa. These results raise the possibility that vortices avoid TBs when we field-cool (cf. \Fig{fig:vortex_decoration}) not only because the superfluid density is enhanced but also because the gap is larger on TBs. This is of course a speculation because the tunneling density of states, which gives the gap, has not been measured on TBs in \PBa.

\section{\label{sec:conclusion}Conclusion}
We have shown that the diamagnetic response is enhanced along stripes that are parallel to TBs in \PBa. These stripes, whose width is on the scale of several $\lambda_{ab}$, repel vortices and act as barriers for their motion. The stripes move at elevated temperatures and disappear when we warm the sample towards the superconducting \Tc. We have ruled out topography as the primary cause of the stripes, as well as the existence of a non-superconducting boundary area between domains.

The stripes that we see exist on a much smaller spatial scale than stripes with similar phenomenology that have been observed in \CoBa\ by Kalisky \& Kirtley \etal\cite{Kalisky2010,Kalisky2011}. Our interpretation, which is based on the direction of the stripes relative to the crystal axes and on the \CoBa\ results\cite{Kalisky2010,Kalisky2011}, is that the stripes are on TBs in the isovalently doped \PBa.

Since TBs are common in underdoped Fe-SCs it is important to understand their properties. This is especially true because of their role in vortex motion, which is one of the most important factors determining the technological utility of superconducting materials \cite{Larbalestier01,Scanlan04,Foltyn2007}. As an example we take measurements of the critical current as a function of doping in \CoBa\cite{Prozorov2009} and in \PBa\cite{Demirdis2013} which show a peak near optimal doping. At first glance this appears to contradict our interpretation of the role of TB as barriers rather than traps for vortices but it is completely consistent -- when TBs create an interwoven mesh\cite{Prozorov2009} of barriers for vortices they can be efficient at preventing vortex motion and thus increase the critical current.

\begin{acknowledgments}
We would like to thank J. E. Hoffman, B. Kalisky, A. Kanigel, A. Keren, D. Podolsky and V. Kogan for discussions, A. Ribak and A. Brenner for help with EDS and EBSD as well as the Micro Nano Fabrication Unit at the Technion. Y.L. acknowledges support from the Technion Russell Berrie Nanotechnology Institute (RBNI). This work was supported by the Israel Science1 Foundation (grant no. 1897/14).
\end{acknowledgments}

\appendix

\renewcommand{\thesection}{\Alph{section}}

\section{\label{app:stripes90} $90^\circ$ stripes}
\begin{figure}
\centering\includegraphics[width=0.95\columnwidth]{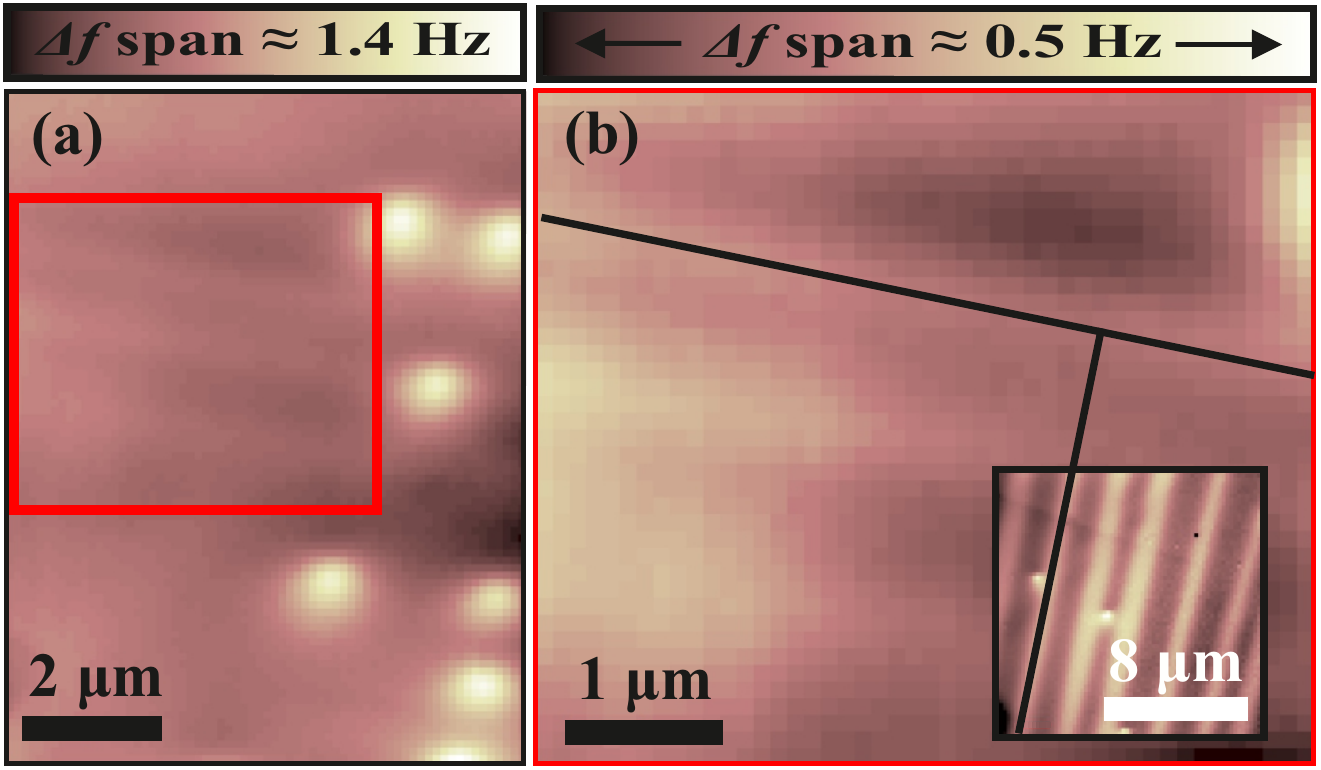}
\caption{Two orthogonal sets of stripes from different areas in the sample. \textbf{(a)} A scan at $T=4.5$~K, $h=280$~nm and $B=1.5$~G. The scan shows  vortices (bright disks) and faint stripes. \textbf{(b)} A zoom on the area marked by the red square in (a) with an expanded color scale to highlight the stripes. A black line marks the direction of the stripes. \textbf{Inset:} The image shown in \Fig{fig:vortex_decoration}(a), with a black line marking the direction of the stripes. As can be seen they are perpendicular to the stripes in the main panel. The $\Delta f$ span for the inset is $2.5$~Hz. }
\label{fig:90deg}
\end{figure}

Most of the stripes we saw were along one direction.  Occasionally we observed stripes at $90^\circ$. An example is in \Fig{fig:90deg}, where we compare the direction of the stripes in \Fig{fig:vortex_decoration}(a) to their direction in another area which is $\approx500$~\um\ away.

\section{\label{app:topo} Topography of the surface near stripes}
We can rule out that the stripes are associated with topographic features higher than several nanometers. This conclusion is based on several facts. The first is that the stripes disappear near \Tc, even when $V_{t-s}$ is a few volts. At such a large $V_{t-s}$ even steps a few nanometers high would be visible because of the strong electrostatic interaction between the tip and the sample. The signal-to-noise ratio of our scans gives upper bounds on the size of topographic features associated with the stripes. If the stripes are associated with wide trenches or bumps their height is no more than $\approx1$~nm. For features much narrower than our spatial resolution the bound on height is inversely proportional to the width. For the scans in this work the resolution is set by the scan height to $\approx100$~nm. With this number we estimate  an upper bound of $20$~nm on the height for $10$~nm wide features and  $4$~nm for $50$~nm wide features.

\begin{figure}
\centering\includegraphics[width=0.75\columnwidth]{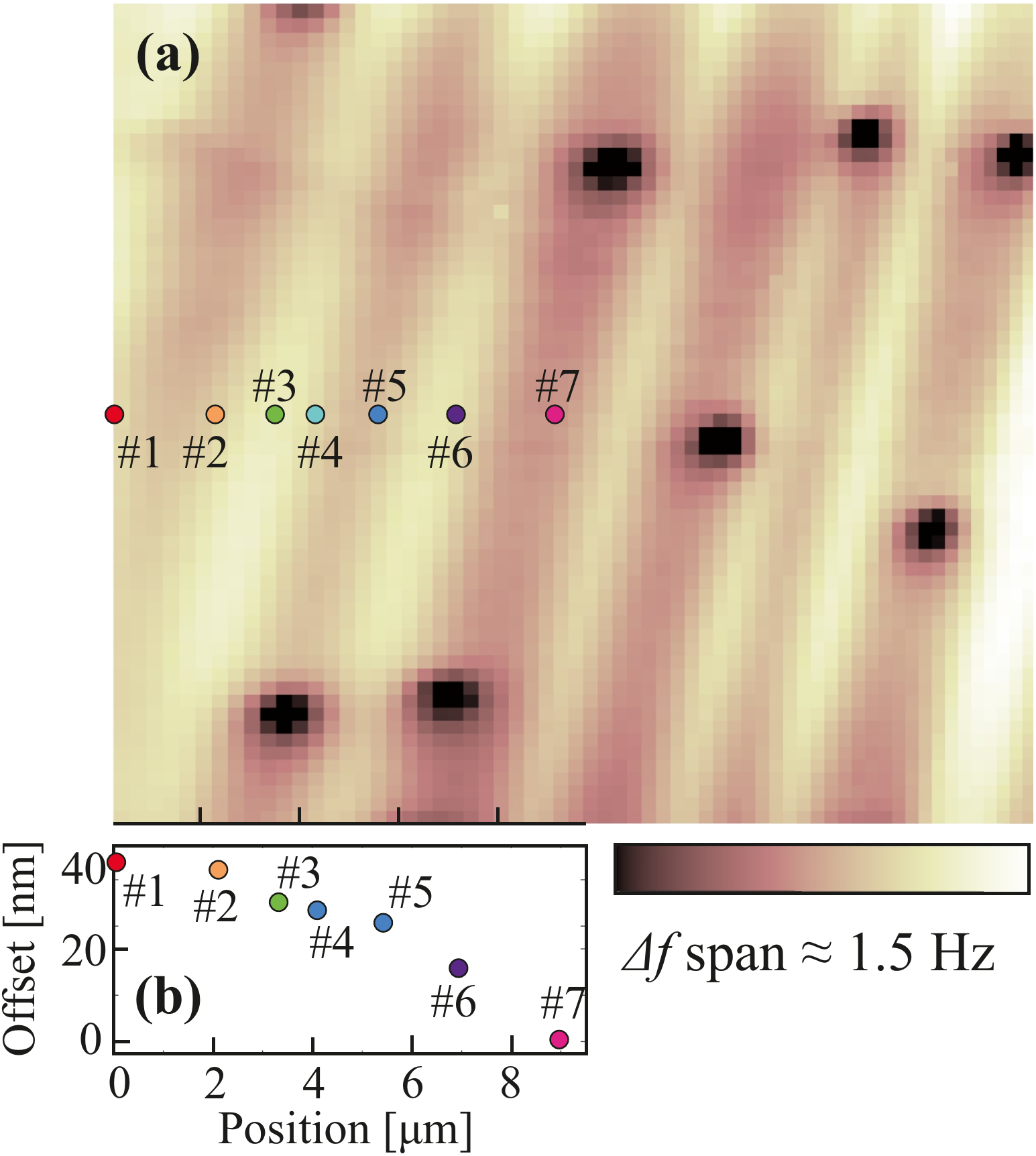}
\caption{Within our accuracy the stripes are not associated with variations of the topography of the sample. \textbf{(a)} Scan (with a plane subtracted) at $T=4.5$~K and $B=-0.5$~G with $h=70$~nm. Also shown are points $\#1-\#7$ where we performed touch-downs. The scale bar for this panel is given by the horizontal axis in (b). \textbf{(b)} Difference in the position of where the tip stops at points $\#1-\#7$. One can see that the variation, which we attribute to creep of the piezoelectric scanner and the tilt of the sample, has no systematic relationship with the stripes. }
\label{fig:topo}
\end{figure}

Additional evidence against height variation as an explanation for the stripes is a series of measurements that show that there are no surface features taller than a few nanometers associated with the stripes. For these measurement we located several conveniently spaced stripes and then brought the tip down to the surface at several points, marked $\#1-\#7$ in \Fig{fig:topo}(a). The resulting curve at each point is called a 'touch-down'. Typically such a curve includes a very sharp drop of the MFM signal that is associated with strong interactions between the tip and the sample\cite{Lamhot2015}. We use the sharp drop to locate the surface. As \Fig{fig:topo}(b) shows when we compare the position of the surface at different points near and on stripes we do not see anything systematic that we can associate with the stripes themselves. The large scale systematic trend that is in \Fig{fig:topo}(b) is the result of creep of our piezoelectric scanner and the tilt of the sample.

\section{\label{app:KK}MFM response for a sheet of reduced $\lambda$}
\begin{figure}[htbp]
\centering\includegraphics[width=0.9\columnwidth]{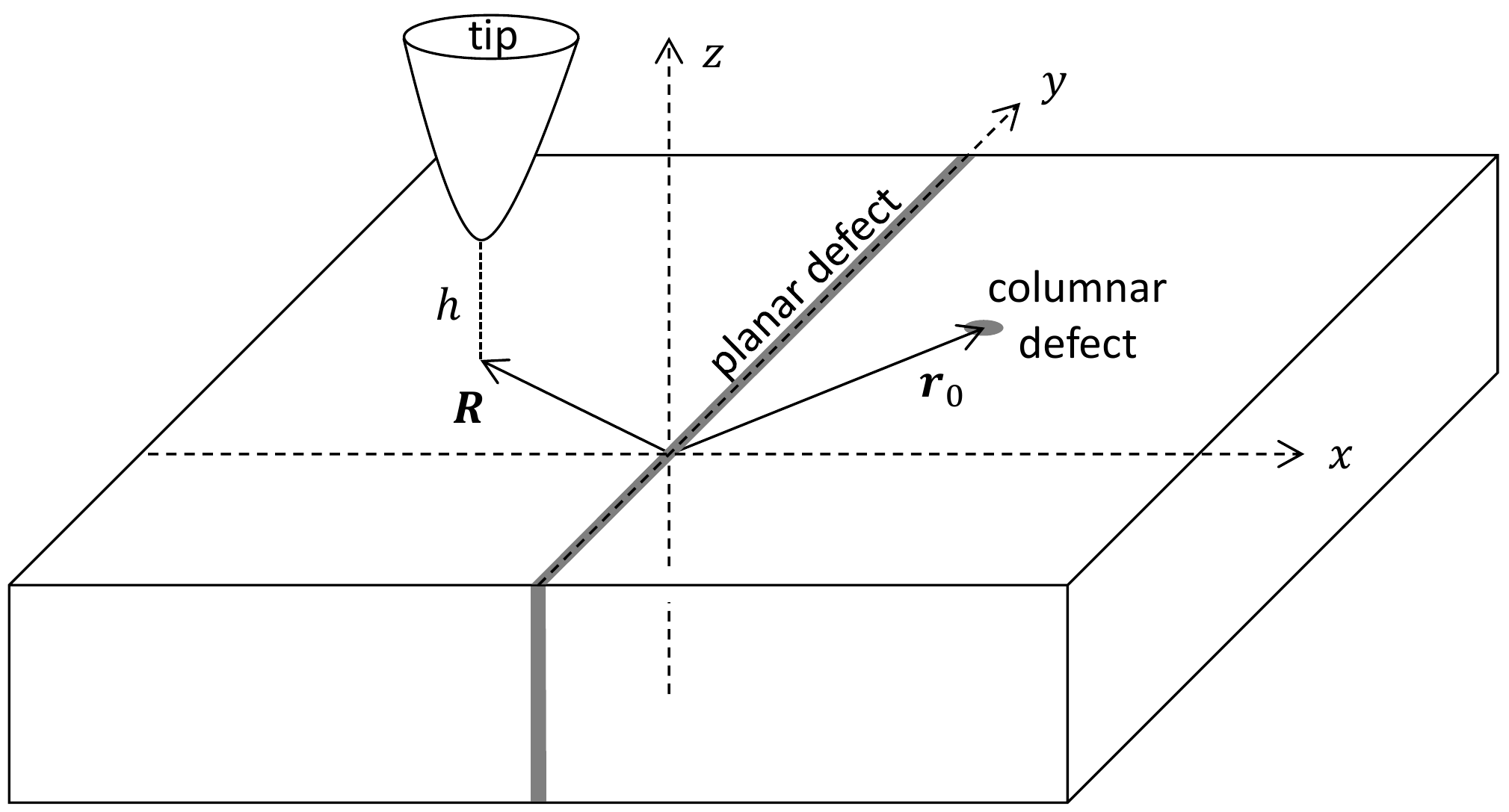}
\caption{Sketch showing the coordinates used in the derivation. The apex of the MFM tip is at $(\vec{R},h)=(X,Y,h)$. The planar defect overlaps the $y-z$ plane and the narrow columnar variation of $\lambda$ is along the $z$-axis at $\vec{r}_0=(x_0,y_0)$.}
\label{fig:sketch}
\end{figure}
Here we closely follow Kogan and Kirtley \cite{Kogan2011} in order to calculate the expected MFM signal near a sheet of enhanced $\rho_s$. In the spirit of Kogan and Kirtley \cite{Kogan2011} we model the enhanced $\rho_s$ by a reduced $\lambda^2$. In order to calculate the MFM signal we need to find the response magnetic field ($\vec{h}^r$) to the magnetic field induced by the MFM tip ($\vec{h}^s$). It is $\vec{h}^r$ that interacts with the MFM tip's magnetization and gives rise to the force which is responsible to the MFM signal. We note that in this Appendix $\vec{h}$ denotes the local magnetic field and $h$ denotes the distance of the MFM tip from the surface.

We choose coordinates so that the superconductor fills the $z<0$ half-space and the MFM tip resides in the $z>0$ half-space (see \Fig{fig:sketch}). To calculate the response to the magnetic field from the tip we must solve the London equations for $z<0$ and the Maxwell equations for $z>0$.

Since there are no currents for $z>0$, it is convenient to define a magnetic scalar potential $\vec{h}^r=\vec{\nabla}\varphi^r$. This potential is defined for $z>0$ and satisfies: $\nabla^{2}\varphi^r\left(\vec{r},z\right)=0$ with $\varphi^r\left(z\rightarrow\infty\right)\rightarrow0 $ and where $\vec{r}\equiv\left(x,y\right)$ are the coordinates parallel to the surface.

Our first task is to calculate $\varphi^r\left(\vec{r},z\right)$ given the magnetic scalar potential of a source $\varphi^{s}\left(\vec{r},z\right)$. More specifically, we want to determine the difference made by a variation of $\lambda^2$: $\psi\left(\vec{r},z\right)=\varphi^r\left(\vec{r},z\right)-\varphi^{r,0}\left(\vec{r},z\right)$. $\varphi^{r,0}\left(\vec{r},z\right)$ and $\varphi^r\left(\vec{r},z\right)$ are the magnetic scalar potentials without and with the variation. For the response it is convenient to use:
\begin{equation}\label{eq:FTphi}
\varphi^r\left(\vec{r},z\right)=\int\frac{d^{2}k}{\left(2\pi\right)^{2}}e^{i\vec{k}\cdot\vec{r}-kz}
\varphi^r_{\vec{k}}
\end{equation}
with $\vec{k}\equiv(k_x,k_y)$, $k\equiv\left|\vec{k}\right|$, and where we assume $z>0$.

\subsection{\label{app:KK.Green}The Green function}
\KK\ solved the problem of determining the response for any source and for any variation of $\lambda^2$ provided that it is the same for all $z<0$ and that it is weak in a perturbation theory sense. For this they calculated the response of a superconductor with a narrow columnar variation of $\lambda^2$ of the form:
\begin{equation}\label{eq:pointdefect}
\lambda^{2}\left(\vec{r}\right)=\lambda_{0}^{2}-\eta^{4}\delta\left(\vec{r}-\vec{r}_{0}\right).
\end{equation}
Here $\lambda_{0}^{2}$ gives the bulk value of $\rho_s^{-1}$, $\eta$ gives the strength of the variation, and $\vec{r}_0\equiv(x_0,y_0)$ is its position (see \Fig{fig:sketch}). Below we will construct a thin plane from a single file of the narrow columns. Consequently, the end result will feature the  length-scale $\beta$ instead of $\eta$, where $\beta^3\equiv\eta^4n$ and $n$ is the density of columnar variations in the single file.

With \Eq{eq:pointdefect} given, \KK\ found that when a magnetic source is at $\left(\vec{R},h\right)\equiv(X,Y,h)$ (see \Fig{fig:sketch}) the leading order correction is:
\begin{eqnarray}\label{eq:PsiKKa}
\psi_{\vec{k}}\left(\vec{r}_{0};\vec{R},h\right)&&=\\
&&\!\!\!\!\!\!\!\!\!\!\!\!\!
\frac{2\eta^{4}}{\lambda_{0}^{2}}
\int\frac{d^2q}{4\pi^{2}}\frac{\left(P-q\right)\vec{q}\cdot\vec{k}  ~e^{i\left(\vec{q}-\vec{k}\right)\cdot\vec{r}_{0}}}
{k\left(p+k\right)\left(P+p\right)}\varphi^{s}_{\vec{q}}\left(\vec{R},h\right).\nonumber
\end{eqnarray}
Here $\vec{q}\equiv(q_x,q_y)$, $q\equiv\left|\vec{q}\right|$, $p^{2}  \equiv  \lambda_{0}^{-2}+k^{2}$ and $P^{2} \equiv \lambda_{0}^{-2}+q^{2}$. $\psi_{\vec{k}}\left(\vec{r}_{0};\vec{R},h\right)$ can be converted back to real space with the transformation in \Eq{eq:FTphi}. $\varphi^{s}_{\vec{q}}\left(\vec{R},h\right)$ in \Eq{eq:PsiKKa} is the magnetic potential of the magnetic tip for $ z < h $:
\begin{equation}\label{eq:FTphisource}
\varphi^{s}\left(\vec{r},z;\vec{R},h\right)=
\int\frac{d^{2}q}{\left(2\pi\right)^{2}}e^{i\vec{q}\cdot\vec{r}+zq}
\varphi^s_{\vec{q}}\left(\vec{R},h\right).
\end{equation}
where $\varphi_{\vec{q}}^s\left(\vec{R},h\right) \equiv  e^{-i\vec{q}\cdot\vec{R}-hq}\varphi^s_{\vec{q}}$ and $\varphi_{\vec{q}}$ is the two-dimensional Fourier transform of the field from the tip on the $z=h$ plane when $\vec{R}=0$.

\subsubsection{\label{app:KK.largeH}Approximation for large height}
Several scales determine the range of $q$ that contributes to the integral in \Eq{eq:PsiKKa}. One is the scale for $p$ and $P$: $\lambda_0^{-1}$. Other scales come from $\varphi^{s}_{\vec{q}}\left(\vec{R},h\right)$. One of these is $h^{-1}$ [see the definition of $\varphi^{s}_{\vec{q}}\left(\vec{R},h\right)$ below \Eq{eq:FTphisource}]. Therefore if $h\gg\lambda_0$ only $k,q\ll\lambda_0^{-1}$ are important and we can replace terms by their small $k,q$ values: $P-q \approx \lambda_{0}^{-1}\exp(-q\lambda_0)$, $p+k \approx \lambda_{0}^{-1}\exp(k\lambda_0)$ and $p+P \approx  2\lambda_{0}^{-1}$. Plugging these approximations into \Eq{eq:PsiKKa} we obtain:
\begin{eqnarray}\label{eq:PsiKK}
\psi_{\vec{k}}\left(\vec{r}_{0};\vec{R},h\right)&&\approx\\
&&\!\!\!\!\!\!\!\!\!\!
\frac{\eta^{4}}{\lambda_{0}}
\int\frac{d^2q}{4\pi^{2}}\frac{e^{-\lambda_0q}\vec{q}\cdot\vec{k}  ~e^{i\left(\vec{q}-\vec{k}\right)\cdot\vec{r}_{0}}}
{ke^{\lambda_0k}}\varphi^{s}_{\vec{q}}\left(\vec{R},h\right)=\nonumber\\
&&\!\!\!\!\!\!\!\!\!\!
\frac{\eta^{4}}{\lambda_{0}}e^{-i\vec{k}\cdot\vec{r}_{0}-\lambda_0k}
\frac{\vec{k}}{k}\cdot\int\frac{d^2q}{4\pi^{2}}\vec{q}  ~e^{i\vec{q}\cdot\vec{r}_{0}-\lambda_0q}
\varphi^{s}_{\vec{q}}\left(\vec{R},h\right).\nonumber
\end{eqnarray}
This can be written as:
\begin{eqnarray}
\psi_{\vec{k}}\left(\vec{r}_{0};\vec{R},h\right)&\approx&\\
&&\!\!\!\!\!\!\!\!\!\!\!\!\!\!\!\!\!\!\!\!\!\!\!\!\!\!\!\!\!\!
\frac{\eta^{4}}{\lambda_{0}}e^{-i\vec{k}\cdot\vec{r}_{0}-\lambda_0k}	\frac{\vec{k}}{ik}\cdot\frac{\partial}{\partial\vec{r}_0}\int\frac{d^2q}{4\pi^{2}}  ~e^{i\vec{q}\cdot\vec{r}_{0}-\lambda_0q}
\varphi^{s}_{\vec{q}}\left(\vec{R},h\right).\nonumber
\end{eqnarray}	
Using \Eq{eq:FTphisource} we see that this given by:
\begin{equation}
\psi_{\vec{k}}\left(\vec{r}_{0};\vec{R},h\right)\approx
\frac{\eta^{4}}{\lambda_{0}}e^{-i\vec{k}\cdot\vec{r}_{0}-\lambda_0k} \frac{\vec{k}}{ik}\cdot\frac{\partial}{\partial\vec{r}_0}\varphi^{s}\left(\vec{r}_0,-\lambda_0;\vec{R},h\right).
\end{equation}
Plugging this result into \Eq{eq:FTphi} we obtain:
\begin{eqnarray*}
\psi_{\vec{R},h}\left(\vec{r},z;\vec{r}_{0}\right) & \approx &\\ &&\!\!\!\!\!\!\!\!\!\!\!\!\!\!\!\!\!\!\!\!\!\!\!\!\!\!\!\!\!\!\!\!\!\!
\frac{\eta^{4}}{\lambda_{0}}\frac{\partial}{\partial\vec{r}_0}
\varphi^{s}\left(\vec{r}_0,-\lambda_0;\vec{R},h\right)\cdot
\int\!\frac{d^2k}{4\pi^2}\frac{\vec{k}}{ik}e^{i\vec{k}\cdot\left(\vec{r}-\vec{r}_0\right)-\left(z+\lambda_0\right)k},\nonumber
\end{eqnarray*}
which gives:
\begin{equation}\label{eq:psipoint}
\psi_{\vec{R},h}\left(\vec{r},z;\vec{r}_{0}\right) \approx
\frac{\eta^{4}}{2\pi\lambda_{0}}\frac{\vec{h}_{\vec{R},h}^{s,||}\left(\vec{r}_0,-\lambda_0\right)\cdot\left(\vec{r}-\vec{r}_0\right)}{\left[\left(\vec{r}-\vec{r}_0\right)^2+\left(z+\lambda_0\right)^2\right]^{3/2}}.
\end{equation}
In the last expression the dot product is between two vectors in the plane, the in-plane part of the source field from the tip evaluated at $\left(\vec{r}_0,-\lambda_0\right)$ and $\left(\vec{r}-\vec{r}_0\right)$.

\subsection{\label{app:KK.MFM}MFM signal}
In order to estimate the MFM signal, we need to calculate $\frac{\partial F_{z}}{\partial h}(\vec{R},h)=-\frac{\partial^2 U}{\partial h^2}(\vec{R},h;\vec{r}_0)$. Here $F_{z}$ is the $\hat{z}$-component of the extra magnetic force exerted on the tip because of the modulation in $\lambda$, and $U(\vec{R},h;\vec{r}_0)$ is the associated potential energy. Therefore:
\begin{eqnarray}\label{eq:U}
	U\left(\vec{R},h;\vec{r}_{0}\right)& = & \\&&\!\!\!\!\!\!\!\!\!\!\!\!\!\!\!\!\!\!\!\!\!\!
	-\frac{1}{2}\int dv'\,\vec{M}(\vec{r}',z')\cdot\vec{\nabla}\psi_{\vec{R},h}\left(\vec{R}+\vec{r}',z'+h;\vec{r}_{0}\right),\nonumber	
\end{eqnarray}
where the dot product is between two three-dimensional vectors. The integration over $\vec{r}',z'$ is over a coordinate system that is centered at the apex of the tip. Every point described by this coordinate system is located at $\left(\vec{r'}+\vec{R},z'+h\right)$ in the global coordinate system we defined in \Fig{fig:sketch}. We start the integration over $z'$ at an arbitrary point $z'= z_0<0$ beneath the tip, remembering that $\vec{M}(\vec{r}',z')$ will always contain a step function that is zero for any $z'<0$.

If we assume the tip is sharp then:
\begin{eqnarray}\label{eq:Usharp}
U\left(\vec{R},h;\vec{r}_{0}\right)& \approx & \\&&\!\!\!\!\!\!\!\!\!\!\!\!\!\!\!\!\!\!\!\!\!\!
-\frac{1}{2}\int\!dz'\vec{\tilde{m}}(z')\cdot\vec{\nabla}\psi_{\vec{R},h}\left(\vec{R},z'+h;\vec{r}_{0}\right),\nonumber
\end{eqnarray}
where we defined a magnetization per unit length $\vec{\tilde{m}}(z')\equiv\int\!d\vec{r}'\vec{M}(\vec{r}',z')$.

\subsubsection{\label{app:KK.mono}Monopole tip approximation}
As a simple example we shall assume that the tip is an infinity long needle with the magnetization pointing along its axis, $\hat{z}$. It will therefore produce a magnetic field of a monopole with an effective magnetization per unit length $\tilde{m}$\cite{Straver08}. The resulting field is:
\begin{equation}\label{eq:monoB}
\vec{h}_{\vec{R},h}^{s}\left(\vec{r},z\right)=-\frac{\mu_{0}\tilde{m}}{4\pi}\frac{\left(\vec{r}-\vec{R}\right)+(z-h)\hat{z}}{\left[\left(\vec{r}-\vec{R}\right)^{2}+\left(z-h\right)^{2}\right]^{3/2}}.
\end{equation}
We now return to \Eq{eq:U} and obtain:
\begin{eqnarray*}
U^\mathrm{mono}(\vec{R},h;\vec{r}_{0}) & = & -\frac{\tilde{m}}{2}\int_{z_{0}}^{\infty}\!\!\!\!\!\!dz'		\Theta(z')\partial_{z}\psi_{\vec{R},h}\left(\vec{R},z'+h;\vec{r}_{0}\right),
\end{eqnarray*}
where $\tilde{m}\equiv\int d\vec{r'}M(\vec{r}',z')$. Integration by parts gives:
\begin{eqnarray*}
U^\mathrm{mono}(\vec{R},h;\vec{r}_{0}) & = & \frac{\tilde{m}}{2}\psi_{\vec{R},h}\left(\vec{R},h;\vec{r}_{0}\right),
\end{eqnarray*}
where we used $\Theta'(z)=\delta(z')$ and the fact that $\psi_{\vec{R},h}\left(\vec{R},z'+h;\vec{r}_{0}\right)$ vanishes for large $z'$. Finally we use the in-plane part of the field from \Eq{eq:monoB} in \Eq{eq:psipoint} and obtain:
\begin{widetext}
\begin{eqnarray*}
U^\mathrm{mono}(\vec{R},h;\vec{r}_{0})  = 
\frac{\tilde{m}}{2}\frac{\eta^{4}}{2\pi\lambda_{0}}
\frac{\vec{h}_{\vec{R},h}^{s}\left(\vec{r}_0,-\lambda_0\right)\cdot
\left(\vec{R}-\vec{r_{0}}\right)}{\left[\left(\vec{R}-\vec{r}_0\right)^2+\left(h+\lambda_0\right)^2\right]^{3/2}}
		 =  \frac{\tilde{m}}{2}\frac{\eta^{4}}{2\pi\lambda_{0}}\frac{\mu_{0}\tilde{m}}{4\pi}
\frac{\left(\vec{R}-\vec{r_{0}}\right)^2}{\left[\left(\vec{R}-\vec{r_{0}}\right)^2+\left(h+\lambda_0\right)^2\right]^{3}}.
\end{eqnarray*}

Now we can calculate the result for a planar defect, $U^\mathrm{mono}_\mathrm{plane}(\vec{R},h)$. Let us assume that the plane is the $y-z$ plane (\Fig{fig:sketch}). With $\vec{R}=(X,Y)$ we find:
\begin{eqnarray}\label{eq:Uplanemono}
U_\mathrm{plane}^\mathrm{mono}(\vec{R},h) &=&\int_{-\infty}^{\infty}\!\!\!\!\!dy_0~n~U^\mathrm{mono}(\vec{R},h;\vec{r}_{0}) = A\tilde{m}^2\beta^3\frac{\pi}{8}\frac{4X^2+(h+\lambda_0)^2}{\left[X^2+(h+\lambda_0)^2\right]^{5/2}},
\end{eqnarray}
where we defined $A\equiv\mu_{0}[(4\pi)^2\lambda_{0}]^{-1}$. Since we are interested in $\partial_{z}F_{z}$ we take two derivatives with respect to $h$ and obtain:
\begin{equation}\label{eq:dFzmono}
\frac{\partial F^\mathrm{mono}_{z}}{\partial h} =- A\tilde{m}^2\beta^3\frac{\pi}{8}\left\{
\frac{12}{\left[(h+\lambda_0)^2+X^2\right]^{5/2}}
+\frac{75X^2}{\left[(h+\lambda_0)^2+X^2\right]^{7/2}}
-\frac{105 X^4}{\left[(h+\lambda_0)^2+X^2\right]^{9/2}}
\right\}.
\end{equation}

\subsubsection{\label{app:KK.trunc}Truncated cone tip approximation}
A more realistic model for an MFM tip than the model in the previous section is the truncated cone approximation \cite{Luan10,Lamhot2015}. We thus assume that our tip is an infinity long cone shaped needle for which the magnetization per unit length is $\tilde{m}_\mathrm{cone}(z)=\tilde{\tilde{m}}(z+h_0)\Theta(z)$ where we defined $\tilde{\tilde{m}}\equiv\Delta\phi t\alpha M_0$ ($\Delta\phi$ is the azimuthal angle of the tip that is magnetically coated, $t$ is the thickness of the coating, $\alpha$ is the cone half-angle, and $M_0$ is the magnetic dipole density of the coating) and $h_0$ is the truncation height. If the magnetic coating is thin and magnetized along $\hat{z}$ the magnetic field in free space is:
\begin{equation}\label{eq:coneB}
\vec{h}_{\vec{R},h}^{s}\left(\vec{r},z\right) = \frac{\tilde{\tilde{m}}h_0}{\tilde{m}}\vec{h}_{\vec{R},h}^{s,\mathrm{mono}}\left(\vec{r},z\right) + \frac{\mu_0}{4\pi}\tilde{\tilde{m}}\left[-\left(1+\frac{z''}{\sqrt{r''^2+z''^2}}\right)\frac{\vec{r}''}{r''^2} +\frac{\hat{z}}{\sqrt{r''^2+z''^2}}\right],
\end{equation}
where $\vec{h}_{\vec{R},h}^{s,\mathrm{mono}}\left(\vec{r},z\right)$ is the source field in \Eq{eq:monoB}. In \Eq{eq:coneB} $\vec{r}''\equiv\vec{r}-\vec{R}$ and $z''\equiv z-h$. Therefore, the in-plane part of the field (to be used in \Eq{eq:psipoint}) is:
\begin{eqnarray*}\label{eq:coneBinplane}
\vec{h}_{\vec{R},h}^{s,||}\left(\vec{r}_0,-\lambda_0\right)=
\frac{\mu_0}{4\pi}\tilde{\tilde{m}} \left[-\left(1-\frac{\lambda_0+h}{\rho}\right)\frac{1}{(\vec{r_0}-\vec{R})^2} -\frac{h_0}{\rho^{3}}\right](\vec{r_0}-\vec{R}).
\end{eqnarray*}
where we defined $\rho^2\equiv\left(\vec{R}-\vec{r_{0}}\right)^2+\left(h+\lambda_0\right)^2$.

Returning to \Eq{eq:Usharp}, we obtain:
\begin{eqnarray}
U^\mathrm{cone}\left(\vec{R},h;\vec{r}_{0}\right)& \approx & -\frac{1}{2}\int_{z_0}^\infty\!dz'\tilde{m}_\mathrm{cone}(z')\partial_{z'}\psi_{\vec{R},h}\left(\vec{R},z'+h;\vec{r}_{0}\right).\nonumber
\end{eqnarray}
This integral can be done by parts. Since $\partial_{z}\tilde{m}_\mathrm{cone}(z)=\tilde{\tilde{m}}\left[(z+h_0)\delta(z)+\Theta(z)\right]=\tilde{\tilde{m}}\left[h_0\delta(z)+\Theta(z)\right]$ we find:
\begin{eqnarray*}
U^\mathrm{cone}\left(\vec{R},h;\vec{r}_{0}\right)& \approx & \frac{\tilde{\tilde{m}}}{2}\left[h_0\psi_{\vec{R},h}\left(\vec{R},h;\vec{r}_{0}\right)+\int_{z_0}^\infty\!dz'\Theta(z')\psi_{\vec{R},h}\left(\vec{R},z'+h;\vec{r}_{0}\right)\right]=
\\&&\!\!\!\!\!\!\!\!\!\!\!\!\!\!\!\!\!\!\!\!\!\!\!\!\!\!\!\!\!\!\!\!\!\!\!\!\!\!\!\!\!\!\!\!
\frac{\eta^{4}}{2\pi\lambda_{0}}\frac{\tilde{\tilde{m}}}{2}\vec{h}_{\vec{R},h}^{s,\parallel}\left(\vec{r}_0,-\lambda_0\right)\cdot\left(\vec{R}-\vec{r}_0\right)\left[\frac{h_0}{\rho^{3}} +\int_{z_0}^\infty\!dz'\frac{\Theta(z')}{\left[\left(\vec{R}-\vec{r}_0\right)^2+\left(z'+h+\lambda_0\right)^2\right]^{3/2}}\right]=
\\&&\!\!\!\!\!\!\!\!\!\!\!\!\!\!\!\!\!\!\!\!\!\!\!\!\!\!\!\!\!\!\!\!\!\!\!\!\!\!\!\!\!\!\!\!
\frac{\eta^{4}\tilde{\tilde{m}}}{4\pi\lambda_{0}}\vec{h}_{\vec{R},h}^{s,\parallel}\left(\vec{r}_0,-\lambda_0\right)\cdot \left(\vec{R}-\vec{r}_0\right)\left[\frac{h_0}{\rho^{3}}+\frac{\rho-(h+\lambda_0)}{\rho\left(\vec{R}-\vec{r}_0\right)^2}\right]
=\frac{\mu_0}{4\pi}\frac{\eta^{4}\tilde{\tilde{m}}^2}{4\pi\lambda_{0}}
\left[\left(1-\frac{\lambda_0+h}{\rho}\right)\left(\vec{R}-\vec{r}_0\right)^{-2}
+\frac{h_0}{\rho^3}\right]^2\left(\vec{R}-\vec{r}_0\right)^{2}.\nonumber
\end{eqnarray*}
Next we integrate over a planar modulation just like we did in order to obtain \Eq{eq:dFzmono}. Along the $yz$ plane we find:
\begin{eqnarray*}
U^\mathrm{cone}_\mathrm{plane}\left(\vec{R},h\right)& \approx &
A\tilde{\tilde{m}}^2\beta^3 \left[
\frac{4}{X}\sin ^{-1}\left(\frac{X}{\sqrt{(h+\lambda_0)^2+X^2}}\right) - \frac{\pi}{\sqrt{(h+\lambda_0)^2+X^2}}\right]\\&&
+A\tilde{\tilde{m}}^2\beta^3h_0 \left[
\frac{4}{(h+\lambda_0)^2+X^2} - \frac{\pi(h+\lambda_0)}{\left[(h+\lambda_0)^2+X^2\right]^{3/2}}\right] +\frac{\left(h_0\tilde{\tilde{m}}\right)^2}{\tilde{m}^2}U^\mathrm{mono}_\mathrm{plane}\left(\vec{R},h\right),
\end{eqnarray*}
where $U^\mathrm{mono}_\mathrm{plane}\left(\vec{R},h\right)$ is defined in \Eq{eq:Uplanemono}. Taking two derivatives with respect to $h$ gives:
\begin{eqnarray}\label{eq:finalmodel}
\frac{\partial F_{z}}{\partial h} & \approx &
-A\tilde{\tilde{m}}^2\beta^3
\left\{\frac{8(h+\lambda_0)}{\left[(h+\lambda_0)^2+X^2\right]^{2}} -\frac{\pi\left[2(h+\lambda_0)^2-X^2\right]}{\left[(h+\lambda_0)^2+X^2\right]^{5/2}}\right\} \\ && -A\tilde{\tilde{m}}^2\beta^3 h_0
\left\{\frac{24(h+\lambda_0)^2-8X^2}{\left[(h+\lambda_0)^2+X^2\right]^{3}} -\frac{\pi(h+\lambda_0)\left[6(h+\lambda_0)^2-9X^2\right]}{\left[(h+\lambda_0)^2+X^2\right]^{7/2}}\right\}+ \frac{\left(h_0\tilde{\tilde{m}}\right)^2}{\tilde{m}^2}\frac{\partial F^\mathrm{mono}_{z}}{\partial h}.\nonumber
\end{eqnarray}
\end{widetext}

\subsection{\label{app:KK.simulation}Implementation and comparison to data}
\begin{figure}
	\centering\includegraphics[width=1\columnwidth]{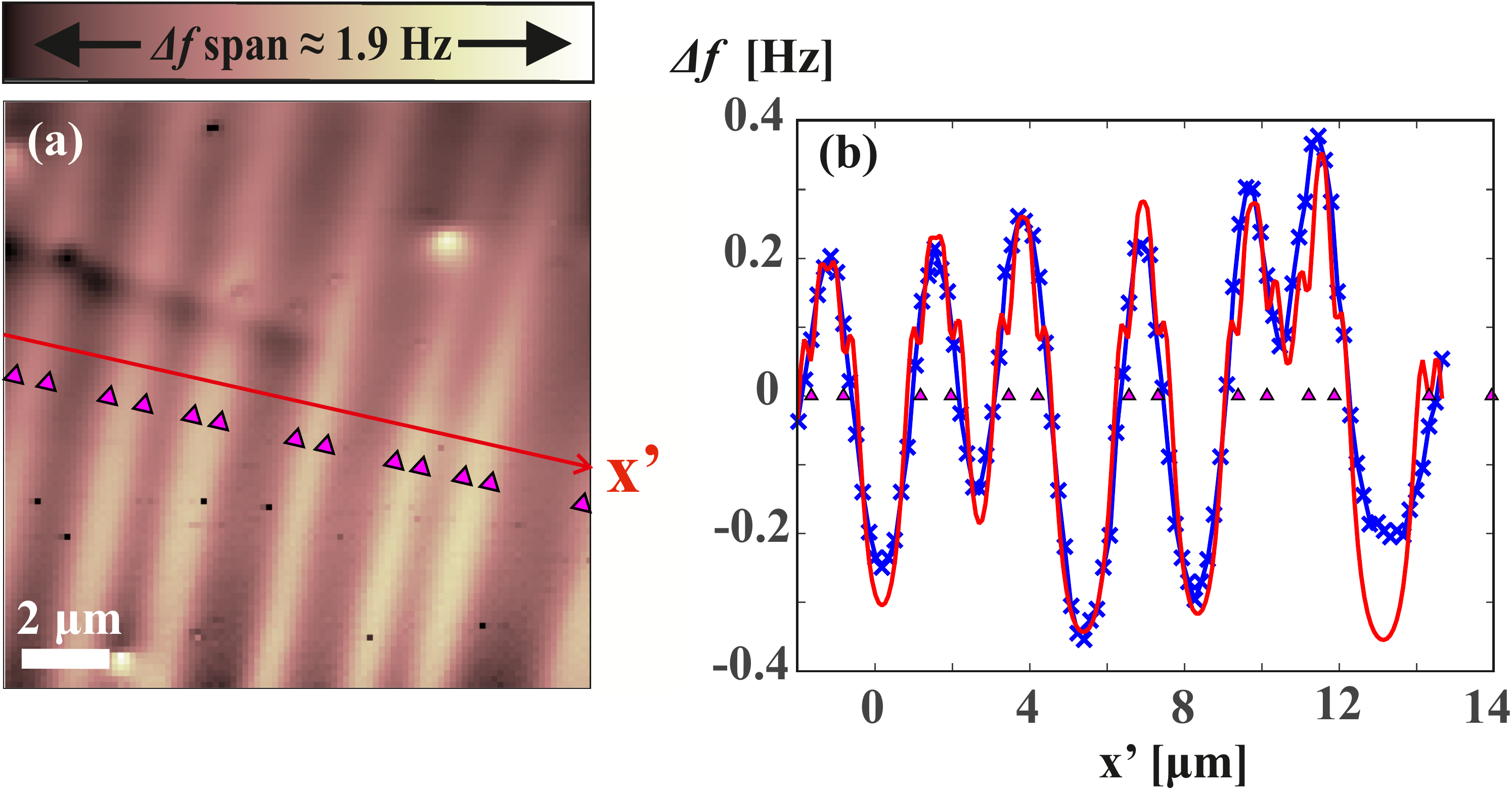}%
	\caption{Comparison between MFM data and \Eq{eq:finalmodel}. \textbf{(a)} MFM scan [the same as \Fig{fig:Tdep}(a)] showing an array of stripes. The triangles show the locations of the TBs that we use in the model and the red arrow shows the line along which we extract data for the comparison with the model. \textbf{(b)} Comparison between data along the red arrow in (a) [blue x marks and line, also shown in \Fig{fig:Tdep}(g)] and the implementation of \Eq{eq:finalmodel} for several stripes (red line). The triangles show the location of the TBs that we used in the model to obtain a curve similar to the data.}
	\label{fig:simulation}
\end{figure}

Here we compare MFM data acquired with a MikroMasch tip\cite{note:Tip} [\Fig{fig:simulation}(a)] with \Eq{eq:finalmodel}. Because of uncertainties in the tip shape we are unable to go beyond a qualitative comparison. To describe the data, we use the simplest version of the truncated cone model \cite{Luan10}. For uniform penetration depth this model gives:
\begin{equation}\label{eq:dFzcone}
\frac{\partial F_{z}^0}{\partial h}=-\frac{\mu_0\tilde{\tilde{m}}^2}{2\pi}
\left[\frac{1}{h+\lambda_0}+\frac{h_0}{\left(h+\lambda_0\right)^2}+
\frac{h_0^2}{2\left(h+\lambda_0\right)^3}\right].
\end{equation}
We set the parameters $\tilde{\tilde{m}}=0.027$~A/m and $h_0=100$~nm to reasonably describe actual touch-down curves\cite{Luan10,Lamhot2015}.

Based on the observations in the main text, we model the narrow stripes in \Fig{fig:simulation}(a) by two TBs each. The different stripe amplitudes can be described by different spacings between the TBs. 
The only other parameters we need in order to emulate the data are the scan height $h=220$~nm, the exact locations of the individual TBs, and the length-scale  $\beta$, which we assume is the same for all stripes.  
We fit for $\beta$ and the stripe location iteratively. First we fit for the TB locations using an initial guess for $\beta$. With the locations fixed, we then fit for a new value of $\beta$ and then use the result to fit for the TB-locations again. The result is in the plot in \Fig{fig:simulation}(b), with the TB locations indicated and $\beta=232$~nm. The emulation mimics our data very well.  Considering all of the uncertainties one should not conclude too much from this agreement but it appears to validate our hypothesis for the internal structure of the stripes, which is below our resolution threshold. 

It is interesting to use the above value for $\beta$ to estimate the strength of the TB-vortex repulsion. This can be done with the aid of Appendix~E in \KK, from which we obtain a scale of $10$~pN/\um\ for a vortex at a distance of $\lambda_0$ from the TB. This matches the scale we obtain from \Fig{fig:histograms}(a).

\begin{thebibliography}{73}
\bibitem [{not({\natexlab{b}})}]{note:fieldcool}%
\BibitemOpen
\href@noop {} {} \bibinfo {note} {This
		means that for every change of the magnetic field we first retracted the tip
		by $\approx30~\mu m$, heated the sample to $T=25~\mathrm{K}>\tc$ for about
		$\approx30$~min and then cooled in the new field to $T<\tc$. After that we
		brought the tip back to the sample and proceeded to scan. In such a procedure
		the scan area can shift by a few microns so where necessary we used
		topographic features to realign. For this we set $V_{t-s}$ to a few
		volts.}\BibitemShut {Stop}%
\bibitem [{not({\natexlab{c}})}]{note:p2p}%
\BibitemOpen
\href@noop {} {} \bibinfo {note} {For
		each $T$ we performed several scans with different $h$. For each
		scan we fitted a plane to obtain an estimate of the height for each data
		point. We then calculated the peak-to-peak amplitudes of the stripes in each
		image and binned the results by height. Since the bins were different at each
		$T$ we interpolated as necessary to compare the amplitude at the same height.
		The result gives the points in \Fig{fig:Tdep}(h) and our estimate for their
		confidence intervals, which are influenced by how far we had to interpolate
		and also reflect the variation from stripe to stripe.}\BibitemShut {Stop}%
\end{thebibliography}
\newcommand{\noopsort}[1]{} \newcommand{\printfirst}[2]{#1}
\newcommand{\singleletter}[1]{#1} \newcommand{\switchargs}[2]{#2#1}
\end{document}